\documentclass[preprint, 12pt, nopreprintline]{elsarticle}

\usepackage{amssymb}
\usepackage{amsmath}
\PassOptionsToPackage{hyphens}{url}
\usepackage{hyperref}
\usepackage{enumitem}

\usepackage{titlesec}
\usepackage{lscape}

\titleformat{\paragraph}
{\normalfont\normalsize\bfseries}{\theparagraph}{1em}{}
\titlespacing*{\paragraph}
{0pt}{3.25ex plus 1ex minus .2ex}{1.5ex plus .2ex}

\usepackage{xcolor}
\usepackage{float}
\newcommand{\na}{N/A}

\begin{document}

\begin{frontmatter}



\title{ARIMA\_PLUS: Large-scale, Accurate, Automatic and Interpretable In-Database Time Series Forecasting and Anomaly Detection in Google BigQuery}


\author{Xi Cheng\corref{cor}}
\ead{chengx@google.com}

\author{Weijie Shen\corref{cor}}
\ead{wjshen@google.com}

\author{Haoming Chen\corref{cor}}
\ead{hmchen@google.com}

\author{Chaoyi Shen} 
\author{Jean Ortega} 
\author{Jiashang Liu} 
\author{Steve Thomas} 
\author{Honglin Zheng} 
\author{Haoyun Wu} 
\author{Yuxiang Li} 
\author{Casey Lichtendahl} 
\author{Jenny Ortiz} 
\author{Gang Liu} 
\author{Haiyang Qi} 
\author{Omid Fatemieh} 
\author{Chris Fry} 
\author{Jing Jing Long}

\cortext[cor]{Corresponding author}
\address{Google}


\begin{abstract}
Time series forecasting and anomaly detection are common tasks for practitioners in industries such as retail, manufacturing, advertising and energy. Two unique challenges stand out: (1) efficiently and accurately forecasting time series or detecting anomalies in large volumes automatically; and (2) ensuring interpretability of results to effectively incorporate business insights. We present ARIMA\_PLUS, a novel framework to overcome these two challenges by a unique combination of (a) accurate and interpretable time series models and (b) scalable and fully managed system infrastructure. The model has a sequential and modular structure to handle different components of the time series, including holiday effects, seasonality, trend, and anomalies, which enables high interpretability of the results. Novel enhancements are made to each module, and a unified framework is established to address both forecasting and anomaly detection tasks simultaneously. In terms of accuracy, its comprehensive benchmark on the 42 public datasets in the Monash forecasting repository shows superior performance over not only well-established statistical alternatives (such as ETS, ARIMA, TBATS, Prophet) but also newer neural network models (such as DeepAR, N-BEATS, PatchTST, TimeMixer). In terms of infrastructure, it is directly built into the query engine of BigQuery in Google Cloud. It uses a simple SQL interface and automates tedious technicalities such as data cleaning and model selection. It automatically scales with managed cloud computational and storage resources, making it possible to forecast 100 million time series using only 1.5 hours with a throughput of more than 18000 time series per second. In terms of interpretability, we present several case studies to demonstrate time series insights it generates and customizability it offers.
\end{abstract}




\end{frontmatter}



\section{Introduction}
\label{intro}

Industrial applications of time series forecasting and anomaly detection have become more challenging with the proliferation of data through digitization. For example, to plan optimally for inventory, retail companies record hundreds of millions of transactions to track and forecast the future demand of millions of products across multiple stores. Similarly, publishers can forecast millions of future impressions and viewership metrics across dozens of regional websites, enabling dynamic inventory allocation to maximize advertising revenue. Two key challenges stand out: scaling up forecasting and anomaly detection capabilities, and ensuring results are interpretable and directly tied to business decisions.

To be more specific, the need for scalability is twofold. First, there is a necessity to automate the forecast process and its maintenance for individuals with limited time series or engineering backgrounds, and for companies with limited resources. Second, there is a need to make forecasting easy and efficient for a large volume of diverse time series, with respect to model training time, resource setup, and management.

Similarly, the need for interpretability involves two key aspects. The first is the ability to understand and quantify the factors that determine forecasted values for diagnosis, customization, and reconciliation with business insights. The second is the need to quantify the inherent uncertainties in data and model predictions to deliver actionable risk assessments for optimal business decision-making.

These requirements make large-scale forecasting and anomaly detection a completely different challenge compared to handling a few time series. Analysts are no longer able to produce thorough model iterations and adjustments by evaluating each time series individually. A new unified framework is required to address automation, speed, interpretation, and uncertainty, and maximize the efficiency of practitioners' involvement to deliver accurate and actionable large-scale forecasts and anomaly detection.

To address the practical challenges inherent in real-world time series forecasting and anomaly detection, we present a novel framework for large-scale, automatic, and interpretable in-database time series forecasting and anomaly detection in Google BigQuery. This framework enables users to:
\begin{enumerate}[label=(\arabic*)]
\item Obtain reliable forecasts and anomaly detection from noisy data without manual intervention: The system directly manages the problems arising from outliers, missing values, abrupt shifts, and gradual changes in time series patterns.
\item Achieve high accuracy across diverse data sets: Rigorous evaluation on public datasets demonstrates superior performance compared to a variety of well-established models, ranking on top of the independently evaluated models across the 42 datasets in the Monash Forecasting archive webpage at the time of writing. Notably, the underlying algorithms are deployed, rigorously tested and actively used for numerous Google internal use cases such as in \cite{TR17}.
\item Benefit from scalability and managed resources: The cloud infrastructure ensures seamless scalability to forecast 100 million time series within a couple of hours, with automated management of underlying hardware (compute and storage) and the models.
\end{enumerate}

This paper focuses on the forecasting methodology, particularly in numerical benchmarking and case studies. We only discuss anomaly detection from the unified modeling perspective in Section \ref{anomaly-detection}. The remainder of this paper is structured as follows: 
\begin{itemize}
\item Section \ref{related-work} introduces related work and its challenges for scalable and interpretable forecast. 
\item Section \ref{model-architecture} discusses the underlying modeling methodology of ARIMA\_PLUS, detailing the algorithms used to handle each component of the time series.
\item Section \ref{system-and-infra} details the infrastructure of ARIMA\_PLUS, encompassing the user interface, model format, and resource management (for both compute and storage) within Google Cloud.
\item Section \ref{experiments} presents performance metrics covering model accuracy and scalability, supplemented by case studies demonstrating interpretability and customization. In particular, we benchmark ARIMA\_PLUS against the 42 public datasets in the Monash forecasting repository \cite{GBW+21}, achieving more accurate forecasting results than a variety of well-established models such as ETS, ARIMA, TBATS, Prophet, DeepAR, N-BEATS, PatchTST and TimeMixer.
\item Section \ref{conclusions} concludes by noting that this innovative framework, featuring a novel, interpretable, and highly accurate modular model architecture and a scalable cloud infrastructure accessible via SQL, establishes a new state-of-the-art for automatic time series forecasting and anomaly detection in industry-scale applications.
\end{itemize}

\section{Related Work}
\label{related-work}
Forecasting models can be categorized broadly into: 
\begin{enumerate}[label=(\arabic*)]
    \item Local, single time series models that include traditional methods like ARIMA, exponential smoothing \cite{McK84} and non-autoregressive models like Prophet \cite{TL18}, and state space models like TBATS \cite{LHR11}. These models are trained individually for each time series in a dataset in order to predict the corresponding time series’ future. 
    \item Global, single time series models like DeepAR \cite{SFGJ20}, TFT \cite{BSNT21}, XGBoost \cite{CG16} and N-BEATS \cite{OCCB19} that are trained globally on many time series but during inference, predict the future of a time series as a function of its own past and other related covariates.
    \item Global, multiple time series models that use the past data of all time series in the dataset to predict the future of all the time series. Such models include the classical VAR model \cite{ZW06} as well as deep learning models like TimesFM \cite{DKSZ23}, to name a few.
\end{enumerate}

Our model, as detailed in the following sections, falls into the first category, a local single time series model. However, instead of jointly fitting the time series parameters, our approach relies on the principle of ``divide-conquer-combine". It allows us to incorporate and compare different methodologies to address challenges of each time series component separately. Therefore, our model is easily extensible to new research advancements, and customizable to the particular use cases, by swapping in better approaches or ensembling multiple approaches at the right module. The model architecture described below in Section \ref{model-architecture} is one instantiation that we find is the most general and therefore productionized in ARIMA\_PLUS.  

The concept of decomposition is well-established in methods such as X13-ARIMA-SEATS \cite{X13} and TRAMO-SEATS \cite{TRAMOS}, though they are primarily used for pure decomposition rather than direct forecasting \cite{HAF00}. While ARIMA\_PLUS shares features like strong interpretability, seasonality/holiday adjustment, and outlier/structural break detection, it offers superior decomposition capability: it can work effectively with irregular seasonal periods, missing data, and high-frequency time series. Furthermore, it incorporates advanced features such as automatic multi-seasonality and multi-holiday detection, as well as trend component model selection via the auto.ARIMA algorithm \cite{RY08}. This article's contribution is to show that our decomposition-based, end-to-end forecasts outperform many time series models that jointly fit all the model parameters.

From a framework perspective, there has been a lot of work to provide time series toolkits, which enable convenient access to various time series models. This includes but is not limited to GluonTS \cite{AKM19} and sktime \cite{LBG19}. Additionally, there are also a few cloud service providers that adopt the existing forecasting packages, e.g., Prophet, ARIMA and Exponential Smoothing are provided in Snowflake Inc's forecasting offerings. Our proposed framework stands apart due to its unique blend of modeling and infrastructure innovations. We developed a new model offering high accuracy and interpretability, and further implemented it directly within the query engine (hence we call it in-database) for highly scalable, fully managed and distributed operation.

\section{Model Architecture}
\label{model-architecture}
In this section, we present a detailed description of the methodology we use to facilitate an automated and interpretable forecast. We first assume the input time series is univariate, i.e., $X_i, Y_i, i=1,\ldots,T$ where $X_i$ indicates the timestamp, $Y_i$ indicates the value, $i$ is the index and $T$ is the index of the last data point used for training. We want to forecast the value of $Y_i$ for a given grid of $X_i > X_T$ for $i=T+1,\ldots,T+H$ where $H$ is the forecast horizon. In the end of the section, the application is generalized to multivariate settings.

\begin{figure}[H]
\centering
\includegraphics[width=0.95\textwidth]{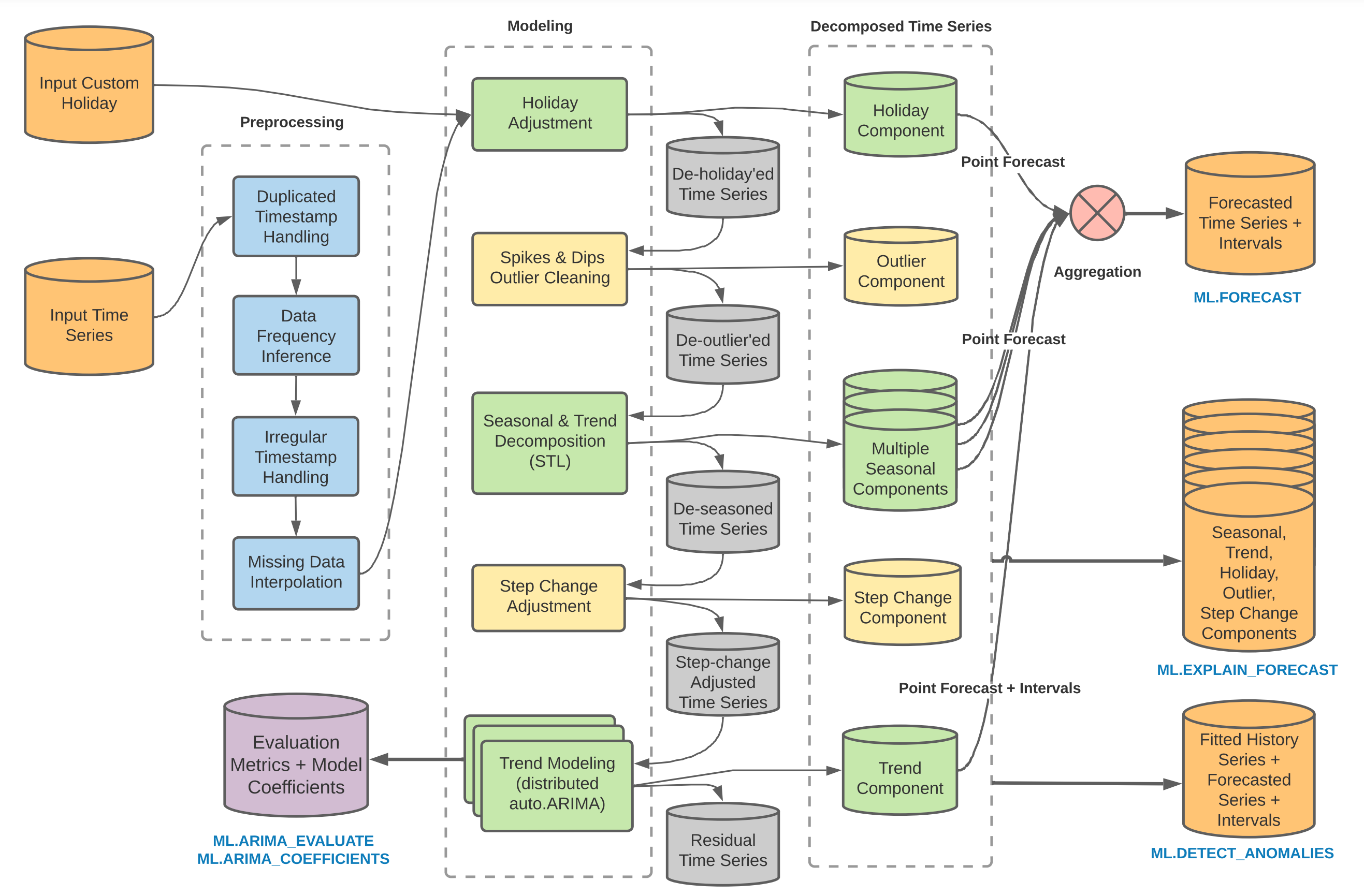}
\caption{The univariate time series modeling pipeline in ARIMA\_PLUS. The ML functions in the diagram are discussed in Section \ref{interface}. 
}\label{fig-univariate}.
\end{figure}

Overall, the forecasting procedure follows this sequential flow of (1) preprocessing, (2) missing value interpolation, (3) holiday extraction, (4) spikes and dips cleaning, (5) seasonality extraction, (6) step change adjustment, (7) trend modeling, (8) seasonality adjustment, and (9) holiday adjustment. One of the unique features of this proposal is that each module of the flow operates on the time series completely independently without relying on other modules, mimicking an assembly line on time series. This sequential modeling feature focuses each step on its own task, which allows it to be iterated, improved or even replaced by other methods for the same task without changes to the other modules, resulting in ultimate controllability over the forecasting behavior. If some module A (e.g., holiday extraction) requires information from another module B (e.g., seasonality extraction), module B (seasonality extraction) is called inside module A (holiday extraction in step 3) and then again in the main flow (in step 5). Such ``divide-conquer-combine" approach helps observe intermediate steps, diagnose each component and interpret the forecasts. We evaluate the performance of the proposed flow sequence and prove its potency compared with other joint modeling methods. However, performance is data-dependent and some internal use cases find using a different sequence works better, which proves the flexibility and customizability of this modular approach. The following sections articulate each module and are organized accordingly. Prediction intervals, covariates, anomaly detection, and other practical forecasting considerations are discussed in addition. The entire univariate time series modeling pipeline is illustrated in Figure \ref{fig-univariate}.

\subsection{Data Processing}
\subsubsection{Time series regularization}
The first step after reading in the data is to standardize it into a typical time series format. Real data often contains various incorrect formats, missing values, and irregularities. This step makes the forecasting procedure robust enough to various kinds of data problems and thus is crucial to the automation of large-scale forecasts over data with potentially heterogeneous variations. Our contribution is to standardize time series treatment between local and absolute timestamps.

First, $X_i$ is formatted into UTC timestamps. ($X_i$, $Y_i$) is removed when $X_i$ is missing or malformatte). Then ($X_i$, $Y_i$) is sorted by $X_i$ and deduplicated. After that, $X_i$ is collected to infer the data frequency by calculating the mode of the distribution of the time intervals, $X_i - X_{i-1}$. Given the inferred frequency, we regularize timestamps into equally spaced grid $X_i^\prime$ and downsample $Y_i$ to $Y_i^\prime$ for the corresponding $X_i^\prime$. 

When the data frequency is close to a calendar frequency (such as monthly or daily), we instead infer a calendar frequency that is potentially not equally spaced in UTC timestamps, such as quarterly. Then we cast timestamps into a month-based indexing system so that they will be equally spaced. We choose month as the unit because it is the most granular typical calendar frequency that contains unequal time intervals. We record time indices of the data as the number of months since the starting date. For example, monthly data of 2021.1.15, 2021.2.15, ..., 2021.12.15, 2022.1.15,... are recorded as 0, 1, ..., 11, 12, ... months since 2021.1.15; quarterly data of 2021.1.1, 2021.4.1, 2021.7.1, 2021.10.1, 2022.1.1,... are recorded as 0, 3, 6, 9, 12 months since 2021.1.1. After forecasts are done, the resulting time indices are mapped back to UTC timestamps.

To deal with Daylight Saving Time (DST), such as a daily time series in America/Los Angeles, we transform input timestamps from UTC timestamps into microseconds since midnight 1960.1.1 at local time. This step may create gaps and duplicates in timestamps because of the daylight saving adjustments. Therefore, we downsample properly using the same method shown above. This way, the forecasting procedure will be operated on a local time system that stays regular through daylight savings. After forecasts are done, local time is transformed back into absolute time where new gaps are interpolated and new duplicates are eliminated. This remedy maps calendar grid to timestamp grid for frequencies lower than hourly. Only the high frequency data during daylight start and end dates require special handling for the gaps and duplicates, which solves the problem for a majority of the use cases.

After data preprocessing, $X_i^\prime$ are regular and equally spaced. ($X_i^\prime$, $Y_i^\prime$) becomes the inputs for the next step. For notation convenience, we always refer to the inputs of each step as ($X_i$, $Y_i$) for the following sections, regardless of the output notation of previous steps.

\subsubsection{Missing Data Interpolation}
Next, $Y_i$ are regularized into standard inputs. We allow three ways to mark missing data. 

\begin{enumerate}[label=(\arabic*)]
    \item Native NAs that are from the inputs or downsampling the inputs;
    \item Specific numbers that are converted to NAs for reasons such as zeros in time series should mean missing data instead of actual zeros;
    \item Specific periods of data are marked as outliers explicitly to be ignored for forecast. 
\end{enumerate}

Then we interpolate the values for the missing data. Suppose the missing index set is $M$ and $Y_i,i\in M$ is missing, we would like to estimate a multidimensional joint conditional distribution of $Y_i, i\in M | Y_j, j\notin M$. 

Traditional interpolation methods include: (1) linear interpolation, (2) splines, (3) Kriging, (4) local linear smoothing such as LOESS \cite{Cleveland1979}. Their choice depends on the nature of the missing data. If the missing data is due to low sampling frequency, linear interpolation is the easiest to carry out. If a degree of smoothness is expected and specified, splines are preferable. If the measurements have little or known uncertainties, Kriging is appropriate. If the uncertainty is large and the missing data is intermittent (spotty), LOESS can effectively reveal the true level. Generally speaking, Kriging has a high accuracy with significantly more computational load, while linear interpolation is simplest and often good enough for a low percentage of missing data in the time series. One of our contributions is making interpolated values also aware of features like seasonality so that the later seasonality adjustment step will not be biased by how we interpolate the missing data.

Interpolated values are typically much smoother than the nearby data points because they represent the estimated conditional mean, which inherently excludes the effect of noise. However, if the missing data make up a substantial portion of the series, especially toward the end, this method runs the risk of under-representing the true variation of the time series. In that case, we can perform an optional follow-up adjustment to back fill correlated noise to the interpolated points. One can use block bootstrap to sample residuals. However, if it is under case 3 where periods of existing data are marked as outliers, it is possible to smooth these outliers locally and capture their high-frequency residuals as noise to be added to interpolated means. This is another contribution of ARIMA\_PLUS, compared to existing approaches.

Suppose the outlier set is $O$ and $Y_i,i\in O$ are the outliers. Their local smoothed values are $\widetilde{Y}_i,i\in O$. $Y_i - \widetilde{Y}_i$ represents the local noise. The interpolated values for these outliers are $\hat{Y}_i,i\in O$. Then the final interpolated values are
$$Y_i-\widetilde{Y}_i+\hat{Y}_i, i\in O.$$

One can interpret it as adding local noise to the interpolate values as in $(Y_i-\widetilde{Y}_i)+\hat{Y}_i$. It can also be interpreted as adjusting the ``average" level of outliers back to the normal range as in $Y_i-(\widetilde{Y}_i-\hat{Y}_i)$.

After the data interpolation step, both $X_i$ and $Y_i$ are regular and ready for modeling.

\subsection{Holidays and events}\label{sec:holiday}
Holiday and event adjustment is the recommended first modeling process, which comes before outlier cleaning and seasonality detection, for the following reasons. First, significant changes may happen to the time series during holidays. To avoid these from being treated as outliers, holiday and event adjustment is done before cleaning. Second, holiday or event effects (in short, holiday effects) are usually on a smaller temporal scale compared to seasonal effects. They usually occur immediately when holidays or events happen and quickly die out when holidays or events end, even when the event lasts for months, such as the Olympics. There are often not much lagging and carrying over effects and thus quite temporary. On the other hand, seasonal patterns can be smoother and more gradual such as summer breaks for students. Holiday effects are often easier to screen out at a local level than seasonal effects, which have to be estimated over longer periods. There are certainly cases such as Thanksgiving and Christmas where holidays combine into a holiday season. It is usually more beneficial to extract holiday effects first to distinguish the effects specific to holiday dates from a general month-long holiday season. We find treating holidays and events first performs well empirically for typical use cases, although some of our internal use cases show that sometimes flipping its order with outlier cleaning and seasonality works better, which demonstrates the flexibility and customizability of this modular approach.

There are existing ways to handle fixed and moving holidays such as in X-13ARIMA-SEATS \cite{X13}. Compared to that, the proposed way below requires much fewer data points to allow a reasonable estimation of holiday effects. Our contributions include (1) being able to capture sudden spikes of holiday effects that do not last the full holiday period; (2) separating and reconciling effects of different holidays on the same date; (3) lightweight holiday detection that does not require multiple years of data; (4) seasonality-aware holiday effects; (5) a default holiday list and the ability to customize; (6) an integrated approach to streamline all holiday steps.

To start modeling, holiday and event dates need to be specified beforehand for both the past and the future. We maintain and publicize a list of big holidays and events that consist of fixed date holidays, floating date holidays and manual events and holidays, ready for use. Meanwhile, users can also provide custom holidays to the system, including the holiday name, holiday date and pre/post number of days that have effect. For a typical single-day holiday such as July 4th in the US, we assume a three day effect window to capture one day before and after. For a long holiday, we curate the start and end of the effect window and that means holidays such as Thanksgiving will cover Black Friday and Cyber Monday. We group holidays by countries and regions so that users only have to input the main geological area the time series is based upon to get a full list of holidays and events that the time series may be subject to. We denote $h\in H_r$ where $h$ is a holiday and $H_r$ is the holiday set for region $r$. The start and end dates are denoted as $S_{th}$ and $E_{th}$, where $t$ is the year, $h$ is the holiday.

The second step is to loop through each holiday to get all the time points that are affected by each holiday. It is a nontrivial step to make it adaptive to different data frequencies because holidays or events are usually defined by the day. For data frequencies finer than daily, we clearly denote all the time points that are affected in each year. For example, for an hourly series, a three-day windowed holiday will have 72 points that are affected every year. In this case, $\chi_{th}=\{X_i|S_{th}\leq X_i\leq E_{th}\}$ is the set of time indices affected by holiday $h$ at year $t$. For data frequencies coarser than daily, we expand the holiday effect window to match the data frequency, such that there is always one point lying within the window each year. For example, holidays will have a window of a week for weekly time series, such that there will always be one point in the weekly time series lying within the holiday windows every year. After the adaptation, we have $\chi_{th}=\{X_i|S_{th}^\prime\leq X_i\leq E_{th}^\prime\}$ where $S_{th}^\prime$ and $E_{th}^\prime$ are modified start and end dates. Since the data is already regularized, the number of time points that are affected is the same across years, i.e. $|\chi_{th}|=N_h$ is the same for any year $t$ for a given holiday $h$. This is the holiday duration in terms of the given data frequency.

We treat each holiday occurrence in the holiday duration to have separate effects and act like ``sub-holidays". For example, for the week-long Thanksgiving on daily time series, it basically factors out to seven separate holidays, i.e. ``1st day of Thanksgiving, 2nd day of Thanksgiving, ..., 7th day of Thanksgiving". This helps us to capture drastically different effect levels across different days of the holidays and tighten the effect window. Here we rewrite $\chi_{th}=\{x_{1th},x_{2th},\ldots,x_{N_hth}\}$, where $x_{jth}$ are ordered timestamps of $\chi_{th}$. We create ``sub-holiday" time series $j$ of holiday $h$ as timestamps $x_{j{t_0}h},x_{j{t_1}h},\ldots,x_{j{t_n}h}$ for year $t_0$ to $t_n$, with their corresponding time series values $y_{j{t_0}h},y_{j{t_1}h},\ldots,y_{j{t_n}h}$.

Next, each ``sub-holiday" gets an estimation of the holiday effect through smoothing. We first union all the time points that are affected by at least one holiday as missing values: $\chi=\bigcup_{th}\chi_{th}$. Then, we interpolate the missing values to create a counterfactual as if there are no holiday effects: $\hat{Y}_i$ for $i$ where $X_i\in\chi$. It is important to recognize potential seasonality effects during interpolation if the holiday estimation is done before seasonality extraction. Then the difference between actual data and the smoothed data is defined as raw holiday effects: $E_i = Y_i-\hat{Y}_i$. We then detect and smooth the holiday effects across multiple years, which allows to further estimate the common holiday effect level from the raw holiday effects for each ``sub-holiday". Suppose $e_{j{t_0}h},e_{j{t_1}h},\ldots,e_{j{t_n}h}$ are the corresponding raw holiday effects $E_i$ for ``sub-holiday" $j$ of holiday $h$ for year $t_0$ to $t_n$. Their smoothed values across years are then written as $\hat{e}_{j{t_0}h},\hat{e}_{j{t_1}h},\ldots,\hat{e}_{j{t_n}h}$. Combining ``sub-holidays", each holiday gets its estimated holiday effect, and we denote them as $\hat{E}_{ih}$ on the original timestamps.  

Finally, some holidays may happen or overlap on the same date. We reconcile colliding holidays following the principle that human behavioral changes during holidays are often due to the time-off and cannot be stacked. For example, during the time-off, one may go on a trip and such lift in expenses can happen on either holiday A or holiday B. However, when holiday A and holiday B are on the same date, such lift will not be stacked or doubled; it should instead be based on the bigger one of the two holidays. Therefore, here we estimate the reconciled holiday effect to be the maximum of all the positive holiday effects plus the minimum of all the negative holiday effects. Suppose holiday $h_1,h_2,\ldots,h_m$ collide on timestamp $X_i$. The reconciled holiday effect on $X_i$ will be $$max\{\hat{E}_{ih_j}|\hat{E}_{ih_j}>0,j=1,\ldots,m\}+min\{\hat{E}_{ih_j}|\hat{E}_{ih_j}<0,j=1,\ldots,m\}.$$

\subsection{Spikes and Dips Outlier Removal}
\label{spikes-and-dips}
A common type of time series anomaly appears as a temporary spike or dip that deviates from the regular pattern. This pattern is easily recognizable, especially on a smooth time series, once the values return to normal levels. However, the task becomes more challenging when the natural variation of the time series is large or when the anomaly is happening at the end of the time series, as it is hard to determine if the anomaly is going to be temporary. Our contribution is to distinguish spikes-and-dips temporary outliers from generic step changes and structural breaks (to be discussed in Section \ref{change-point}) in a reliable and scalable way.

Once holiday effects are removed, an adaptive double exponential smoothing is applied to the time series in two ways: from front to back and from back to front. Suppose the data and trend smoothing parameters are $\alpha$ and $\beta$. It follows the typical formulation: 

\begin{eqnarray}
\textrm{forward: }l_i = \alpha Y_i + (1-\alpha)(l_{i-1}+b_{i-1});b_i = \beta(l_{i} - l_{i-1})+(1-\beta)b_{i-1}\\
\textrm{backward: }l_i^\prime = \alpha Y_i + (1-\alpha)(l_{i+1}^\prime+b_{i+1}^\prime);b_i^\prime = \beta(l_{i}^\prime - l_{i+1}^\prime)+(1-\beta)b_{i+1}^\prime,
\end{eqnarray}

where $l_i$ and $b_i$ are the smoothed level and slope from forward smoothing at time $i$, while $l_i^\prime$ and $b_i^\prime$ are the smoothed level and slope from backward smoothing at time $i$.

The smoothing parameter $\alpha$ is relatively large to adapt to seasonality and $\beta$ is relatively small to be robust against local variations. For each direction, we calculate the difference between the original time series and the smoothed time series for each point and their robust z scores, which can be interpreted as a degree of surprise. For example, $d_i = Y_i - l_i$, $z_i=d_i/std(d_i)$, where $std(d_i)$ is the standard deviation of $d_i$. Then, $A_{forward}=\{i||z_i|>q\}$ are all the indices whose z scores ($z$) are above a user specified  threshold $q$ from the forward smoothing. Similarly, we can get $A_{backward}=\{i||Y_i - l_i^\prime|/std(Y_i - l_i^\prime)>q\}$.

Only points that have significantly large z scores from both directions are marked as spikes and dips. Furthermore, only those that have the same signs in the differences are selected to distinguish spikes and dips from step changes: $A_{both}=A_{forward}\bigcap A_{backward}\setminus \{i|(Y_i-l_i)(Y_i-l_i^\prime)<0\}$. However, because some burn-in is needed for double exponential smoothing, we use one-way detection on both edges of the time series of length $L$. Furthermore, spikes and dips that are seasonal and repeated are kept in the original time series for the following seasonality module to pick up. Here, $\{i|i<L\}$ means indices of first $L$ points of the time series, while $\{i|i>T-L\}$ are those of last $L$ points. Finally, we have:
\begin{eqnarray}
A_{detected}=A_{both}&\bigcup& A_{forward}\bigcap \{i|i>T-L\}\nonumber\\
&\bigcup& A_{backward}\bigcap \{i|i<L\}\nonumber\\
&\setminus&\{i|i\textrm{ exists every year}\}
\end{eqnarray}

The detected spikes and dips are marked as missing values. The interpolation module as mentioned earlier is used to estimate what the values they should take as if they were not anomalies. Due to the detection methodology, spikes and dips typically manifest as isolated events or exhibit only brief durations; consequently, they are often easily interpolated. The difference between the actual values and the estimated values are extracted as the spikes-and-dips component of the time series, and are removed from the inputs. Only very large, obvious anomalies that are greater than seasonality are removed in this step. If further anomaly detection is needed, this step can be called again on deseasoned time series after the next step which is seasonality detection.

\subsection{Seasonality}
Seasonality is a repeated, calendar-based pattern in a time series. For example, when dealing with retail sales, it is likely the case that different days of the week have a strong effect on overall sales (weekly seasonality, such as weekday vs. weekend). Additionally, the time of the year also has a clear effect on retail sales (yearly seasonality). Compared with holidays and events, seasonality results in effects that are typically more gradual and smooth. As it is one of the most noticeable patterns in time series forecasting, it is extremely important to model the seasonality correctly. An underestimated or misaligned seasonal effect is often immediately apparent upon visual inspection.

In order to extract the seasonal component from the data and model multiple seasonality frequencies (e.g., daily, weekly and yearly), we need to extract the trend component of the time series jointly. One popular iterative procedure is Seasonal and Trend decomposition by Loess (STL) \cite{CC90}, which smooths the time series to filter out high and low frequency signals. Here we enhance STL in terms of its treatment towards the following practical issues: (1) multiple seasonal periods, (2) calendar and irregular periods, (3) seasonality detection, (4) trendy or dynamic seasonalities (e.g., a growing day of week effect) and (5) seasonal patterns (e.g., larger day of week effect in summer than in winter).

Although some enhancements, such as multiple seasonal periods \cite{BHB21} and seasonality detection methods (e.g., Kruskal-Wallis, Canova-Hansen, periodogram, and ACF), are available, other crucial issues mentioned previously are rarely discussed in the literature. Because of these unique improvements in our approach, it has better performance in complicated situations compared to sinusoidal regression or Fourier series type methods. In more detail,
\begin{enumerate}[label=(\arabic*)]
    \item It can capture the day-of-month effect precisely, such as the beginning of month or end of month spikes (typical in accounting time series) despite each month having a different number of days.
    \item It can exclude overspecified periods that are not statistically significant given the presence of the other seasonalities.
    \item It can fit very spiky seasonalities without the need of extra terms.
    \item It can capture a yearly pattern on the day-of-week effect, accurately predicting the cycle of weekend / weekday lift.
    \item It can fit and extrapolate a changing seasonal pattern instead of treating it as fixed throughout history.
    \item It can deal with seasonality whose period is a multiple of that of another such as daily vs weekly and correctly attributes to the higher frequency.
\end{enumerate}

Here we briefly discuss how irregular seasonal periods and trendy seasonal seasonalities are fitted.

Seasonal spikes happening at the start or the end of month on a daily time series are common for data like those in finance and accounting. The seasonal period is monthly, but the exact intervals range from 28 to 31 days. Neither STL nor Fourier series can exactly handle dynamic periods, leading to damped effects or inaccurate dates of effects. The approach we use here is to standardize the number of points within each month and use interpolation to translate values between calendar months and standardized months. Suppose the original time series has timestamps $(X_1,\ldots,X_T)$ within a month, where $T$ range from 28 to 31 for daily time series and from $28\times24$ to $31\times24$ for hourly time series. We create a standardized grid $(X_1^\prime,\ldots,X_{T^\prime}^\prime)$, where $X_i^\prime$ stands for $i/30$th of a month from the start of the month. $T^\prime$ will be 30 for daily time series and $30\times24$ for hourly time series. We interpolate the original $(Y_1,\ldots,Y_T)$ to the new grid $(Y_1^\prime,\ldots,Y_{T^\prime}^\prime)$, conduct STL and get seasonal component $(S_1^\prime,\ldots,S_{T^\prime}^\prime)$, interpolate back to original timestamps $(X_1,\ldots,X_T)$ as $(S_1,\ldots,S_T)$, and finally the deseasoned time series is $(Y_1-S_1,\ldots,Y_T-S_T)$.

Two major reasons we prefer STL to Fourier series are (1) STL has a robust option for outliers, (2) STL has a seasonal smoothing window parameter to control how seasonality changes every cycle. A large STL seasonal window forces a constant seasonality while a short one allows seasonality to drift from cycle to cycle. We have found that in practice, there are two major sources of cycle over cycle changes for weekly seasonality in particular: (1) a trend component where, for example, the size of the Monday lift to the Sunday drop becomes larger and larger, contributing to an increasing weekly effect through years, (2) a seasonal component where, for example, the size of Monday lift is larger during winter than during summer. We call them trend over seasonality and seasonality over seasonality, respectively. Using a relatively small seasonal smoothing window and a degree of one will capture some trend and seasonality in the seasonal component when decomposing the time series. However, the extrapolation of the seasonal component itself becomes a forecasting problem that is not as simple as that when having fixed seasonal effects. Here, we use a technique called ``forecasting by leaping" to break down a seasonal component into sub-time series. For example, for a weekly seasonality on a daily time series, we create seven sub-time series from the weekly seasonality, i.e. Monday time series, Tuesday time series, ..., Sunday time series. Each sub-time series is extrapolated to the future via a trend model such as double exponential smoothing before combining back together. Suppose $(S_1,\ldots,S_{T})$ has a period of $K$, we create sub-time series $(S_k, S_{k+K},\ldots, S_{k+\lfloor T/K\rfloor K})$, where $k=0,\ldots,K-1$. The extrapolated time series is $(S_{k+\lfloor (T/K\rfloor+1) K},\ldots,S_{k+\lfloor (T+H)/K\rfloor K})$ which is forecasted independently before putting back to $(S_{T+1},\ldots,S_{T+H})$.

To handle multiple seasonalities, a default set of periods is specified based on the input data frequency. For example, typical seasonalities considered for daily time series are weekly, monthly, quarterly, and yearly seasonalities. These periods will be individually and iteratively fitted, extracted, detected, deduplicated, extrapolated to the future and recorded as the seasonal components of the time series, each representing one period (e.g., weekly, yearly). The difference between the actual values and the sum of the seasonal components of the time series is passed to the next module, which represents deseasonalized data.

\subsection{Change point detection and step change adjustment}
\label{change-point}
Another less common but more influential type of anomaly in the time series is a change point such as structural breaks, level shifts, or trend drifts. These cases occur when the levels or trend shift significantly across a time point, thereby breaking the stationarity assumption relied upon by many time series models. Common causes of these shifts include changes in key metrics, the introduction of new products, system migrations, and unpredictable ``black swan" events, such as the impact of COVID-19. The existence of such a change point often suggests a change of the underlying time series structure, which could jeopardize the accuracy of forecast. The change point detection and step change adjustment step attempts to identify and neutralize the change, yielding a cleaned time series that represents the data after removing the detected shifts. It is composed of two sequential steps: change point detection and step change adjustment. Our contribution to the change point detection, compared to traditional methods, is to treat it as generic tests over out-of-sample errors with recursive forecasts, and to use smart change period combinations to decide the number of change points in the time series. Our contribution to the step change adjustment is to streamline a Chow's test-type adjustment \cite{C60} and remove the changes.

The change point detection step is based on the idea of creating rolling local forecasts and testing if the residuals (i.e., $|actuals - forecasts|$) of the next $m$ points are significantly different from a baseline level observed in the history. It follows the following steps:
\begin{enumerate}[label=(\arabic*)]
    \item On a time series $Y_0, \ldots, Y_i, \ldots, Y_n$, a local forecast $F_0^i, F_1^i, \ldots, F_m^i$ is made at the index $i$ using historical values $Y_0, \ldots, Y_{i-1}$.
    \item The residuals can be written as $R_j^i=F_j^i-Y_{i+j}$ for $j=0,\ldots,m$. Let $R^i=\sum_jR_j^i$ be the total residuals and $z^i$ be their respective robust z scores over $i=0,\ldots,n$.
    \item The time points that have big total residuals of the next $m$ points are labeled as \textit{Forward Potential Change Starts}(FPCS). $T_{FPCS}=\{i||z^i|>z\}$ for a given $z$ threshold. 
    \item Then we reverse the time series $Y_i^\prime=Y_{n-i}$ and create backward rolling forecasts $F_0^{\prime i}, F_1^{\prime i}, \ldots, F_m^{\prime i}$.
    \item We label time points that have big residuals of next $m$ points (these will be $m$ points prior to $i$ because the time series has been reversed) as \textit{Backward Potential Change Ends}(BPCE). $T_{BPCE}=\{n-i||z^{\prime i}|>z\}$, where $z^{\prime i}$ are robust z scores of residual sums $R^{\prime i}=\sum_jR_j^{\prime i}=\sum_j(F_j^{\prime i}-Y_{i+j}^\prime)$.
    \item Then we pair up $T_{FPCS}$ and $T_{BPCE}$ into [\textit{Potential Change Period Start}, \textit{Potential Change Period End}]. For example, if $i\in T_{FPCS}$, $j\in T_{BPCE}$ and $0<j-i<m$, then $[i,j]$ forms a pair. If there is no $j\in T_{BPCE}, 0<j-i<m$ for a given $i\in T_{FPCS}$, $[i, i+m]$ forms a pair. If there is no $i\in T_{FPCS}, 0<j-i<m$ for a given $j\in T_{BPCE}$, $[j-m, j]$ forms a pair. In addition, consecutive FPCS and BPCE are merged into a larger window. 
    \item Finally, overlapping and short change periods are merged, expanded and de-duplicated to construct the final sets of [\textit{Change Period Start}, \textit{Change Period End}].
\end{enumerate}

\begin{figure}[h!]
        \centering
        \includegraphics[width=0.9\linewidth]{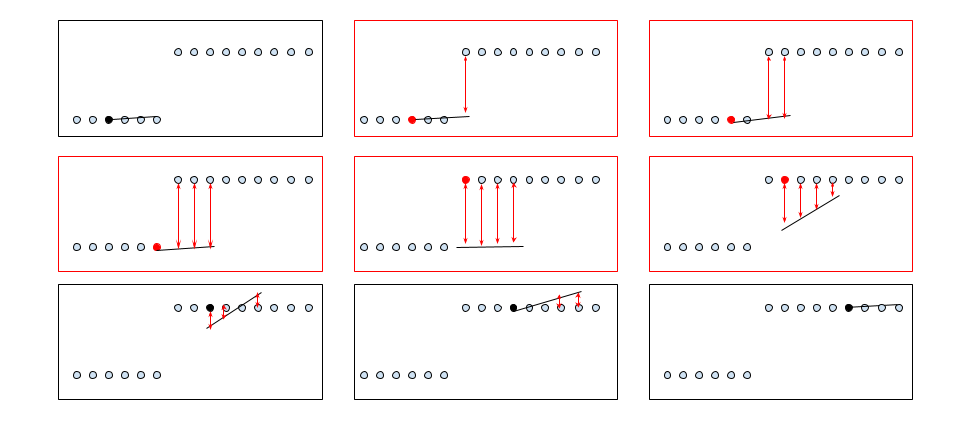}
        \caption{Illustration of change point detection in the forward step. A total of 9 local forecasts are shown. Black lines stand for local forecasts $F^i$. Red arrows stand for large residuals $R_j^i$. Red boxes stand for the 5 local forecasts whose total residuals $R^i$ are above the threshold. Finally, red dots stand for the Forward Potential Change Starts (FPCS).}
        \label{fig:change_detection}
    \end{figure}

Figure \ref{fig:change_detection} is an illustration of how the change point detection works for steps 1-3. 

Once ``Change Period Starts" and ``Change Period Ends" are automatically identified, the second step is to test if there are any level and trend shifts before and after the change. Suppose we only have one change period [\textit{Change Period Start}, \textit{Change Period End}], the process is executed in the following steps:
\begin{enumerate}[label=(\arabic*)]
    \item First, the values within the change window [\textit{Change Period Start}, \textit{Change Period End}] will be set to NAs.
    \item Next, we label the pre-change stable period as (\textit{Series Start}, \textit{Change Period Start}) and the post-change stable period as (\textit{Change Period End}, \textit{Series End}).
    \item Then a Chow test is conducted to test if there is a structural break assuming pre-change stable period follows a different trend from post-change stable period.
    \item The intercept and slope differences between the two linear estimations are put into a change adjustment time series so that $cleaned\_time\_series  = original\_time\_series - change\_adjustment\_time\_series$. The change adjustment will make sure minimal changes are made to the end of the time series, and will only adjust historical stable periods so that they align with recent data. 
    \item Finally, the change periods will be marked as the missing data and interpolated from the adjusted stable periods. When the change periods are large, seasonal and noise-aware interpolations are conducted to avoid biasing the seasonality and the noise level of the original time series. The cleaned time series will then be free from level and trend changes and close to the recent data.
\end{enumerate}

When we have multiple change windows, stable periods are defined as periods in between the change periods such as (\textit{Previous Change Period End}, \textit{Next Change Period Start}), with the first and last stable periods being (\textit{Series Start}, \textit{First Change Period Start}) and (\textit{Last Change Period End}, \textit{Series End}). The stable periods are illustrated in Figure \ref{fig:change_adjustment} below.
    \begin{figure}[h!]
        \centering
        \includegraphics[width=0.9\linewidth]{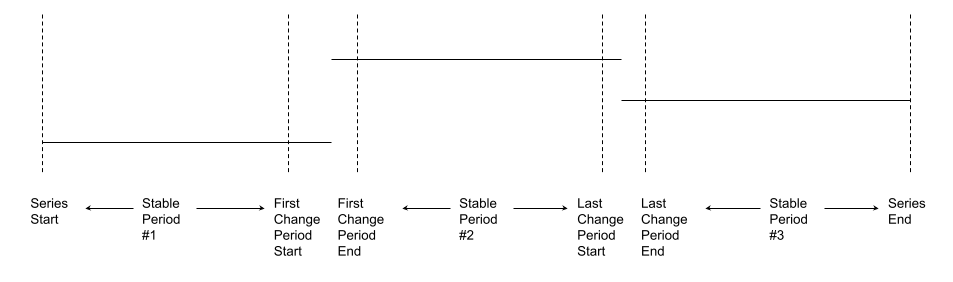}
        \caption{Stable periods for step change adjustment.}
        \label{fig:change_adjustment}
    \end{figure}

\subsection{The trend model}
Following all the previous processes, the time series is left with only the trend component and the residual component.

For trend modeling, we choose the ARIMA model due to its mature foundation and good practice performance. For ARIMA models, the parameters available are the orders ($p$, $d$, $q$) , where they represent the number of autoregressive terms, the order of differences, and the number of moving average coefficients respectively. 

Although $\text{ARIMA}$ models are powerful, a significant barrier to their widespread use in forecasting is that the necessary order selection process is typically regarded as subjective and complex to implement. The auto.ARIMA algorithm \cite{RARS02} solves this problem and works as follows:
\begin{enumerate}[label=(\arabic*)]
    \item The order of differences $d$ is determined using repeated KPSS tests.
    \item The time series data is differenced $d$ times to become stationary.
    \item The values of $p$ and $q$, and whether the ARIMA model should include a linear drift term are then determined by minimising the AIC of the ARIMA model. Note that the drift term is applicable when d is determined to be 1. 
    \item For each \{$p$, $q$\} pair and optionally the linear drift term, the ARIMA model is trained to optimize the associated model parameters, namely autoregression coefficients and moving average coefficients. The fitting process is performed via the standard maximum log likelihood technique and optimized via the L-BFGS algorithm \cite{RPJ95}. 
\end{enumerate}

After the objective function converges, we can obtain the variance, which is the sum of squares calculation between the fitted time series and the input time series. The variance is used for estimating the uncertainty of the time series, and therefore the prediction interval later. 

The search space is controlled by a maximum ARIMA order, which is the sum of $p$ and $q$. As the maximum ARIMA order values increase, not only do the number of candidate models increase, but also the complexity. This causes the training time of the model to increase significantly as the maximum ARIMA order increases. In Section \ref{evaluate-hyperparameters}, we quantify the effect of the maximum ARIMA order on both forecasting accuracy and training time, which allows us to determine an optimal default value for this parameter. This optimization constitutes a key contribution of this paper.

It’s worth noting that there is also a seasonal ARIMA model which can model both seasonality and trend together. However, in practice the seasonal fitting is very computationally expensive and memory-intensive. Therefore, we only apply the non-seasonal ARIMA model for trend models, while leaving the seasonal component to be handled by the STL algorithm as mentioned earlier. 

\subsection{Point forecast and Prediction intervals}
During forecasting, we compute both point forecast and prediction interval. For point forecasts, we extrapolate the fitted components of trend, seasonality and holiday effects if they are present in the given time series, and then sum them up to form the forecasts of the input time series. 

Beyond point forecasts, quantifying uncertainty through prediction intervals is often crucial. These intervals provide a range of plausible future values, conveying the confidence associated with the forecast. Once the mean of non-trend components, such as seasonality and holiday effects, is removed, the uncertainties and modeling errors from these components become bundled with the trend variations and uncertainty. Given a specified confidence level, corresponding prediction upper and lower bounds can be derived from the residual part of the trend model. The resulting prediction intervals therefore contain model uncertainties from all components. 

To illustrate, for a 95\% confidence level, the prediction interval for $y_t$ is calculated as $(y_t-1.96\frac{\sigma}{n}, y_t+1.96\frac{\sigma}{n})$, where 1.96 is the corresponding z-score for 95\% confidence level, $n$ is the size (number of observations) of the sample, and $\sigma$ is the standard error, or the square root of the variance estimated from the ARIMA fitting process. The variance is further projected into the future via the ARIMA process. For a confidence level that is not 95\%, we use the inverse cumulative distribution function (CDF) to get the z-score multiplier. This facilitates the construction of confidence intervals for arbitrary confidence levels as specified by the user.

\subsection{Covariates}
So far, the discussion is all about univariate time series where the forecasted values depend merely on historical data. In many cases, the forecasted values are also affected by some covariates (also known as external factors or regressors), which form the multivariate time series. We adopt the methodology of “regression with ARIMA errors” in \cite{RY08} to model the external regressors.

In our algorithm, we extend the above univariate modeling to multivariate by using the algorithm flow below. For the target value $y$ and its covariates $x_1, \ldots, x_n$, the algorithm is as follows:
\begin{enumerate}[label=(\arabic*)]
    \item Solve the following ridge regression to estimate $\hat{\beta}$: $y_t=c +\beta _0t+\beta _1x_{1,t}+\ldots+\beta _nx_{n,t}+\epsilon _t$ on historical data with optional $L2$ regularization. The timestamp term $t$ is to model the linear trend. We solve it using the closed-form solution of ridge regression.
    \item Calculate the fitted part $\hat{y_t}=\hat{\beta _1}x_{1,t}+\hat{\beta _2}x_{2,t}+\ldots+\hat{\beta _n}x_{n,t}$.
    \item Get residual part $r_t=y_t-\hat{y_t}$. Apply univariate ARIMA (and auto.ARIMA) to get $r_{future}$, where $future$ contains all the future time points.
\end{enumerate}

The forecasting process is as follows:
\begin{enumerate}[label=(\arabic*)]
    \item Given a future input covariate $x_{future}$, calculate the forecasted value $\hat{y}_{future} = \hat{\beta} x _{future}$ from the linear model. 
    \item Get the future $r_{future}$ from the univariate forecasting process.
    \item The final point forecast is $y_{future}=r_{future}+\hat{y}_{future}$. The prediction interval is still based on the variance of the trend modeling via ARIMA. 
\end{enumerate}

For the covariate features in multivariate time series forecasting, we also apply automatic dummy encoding on the categorical data. This process transforms each category into a binary variable (0 or 1), enabling the model to learn the distinct impact of each category. This automatic conversion simplifies data preparation for users by allowing the model to directly incorporate categorical features, which is a contribution of our proposed model. 

\begin{figure}[H]
\centering
\includegraphics[width=0.95\textwidth]{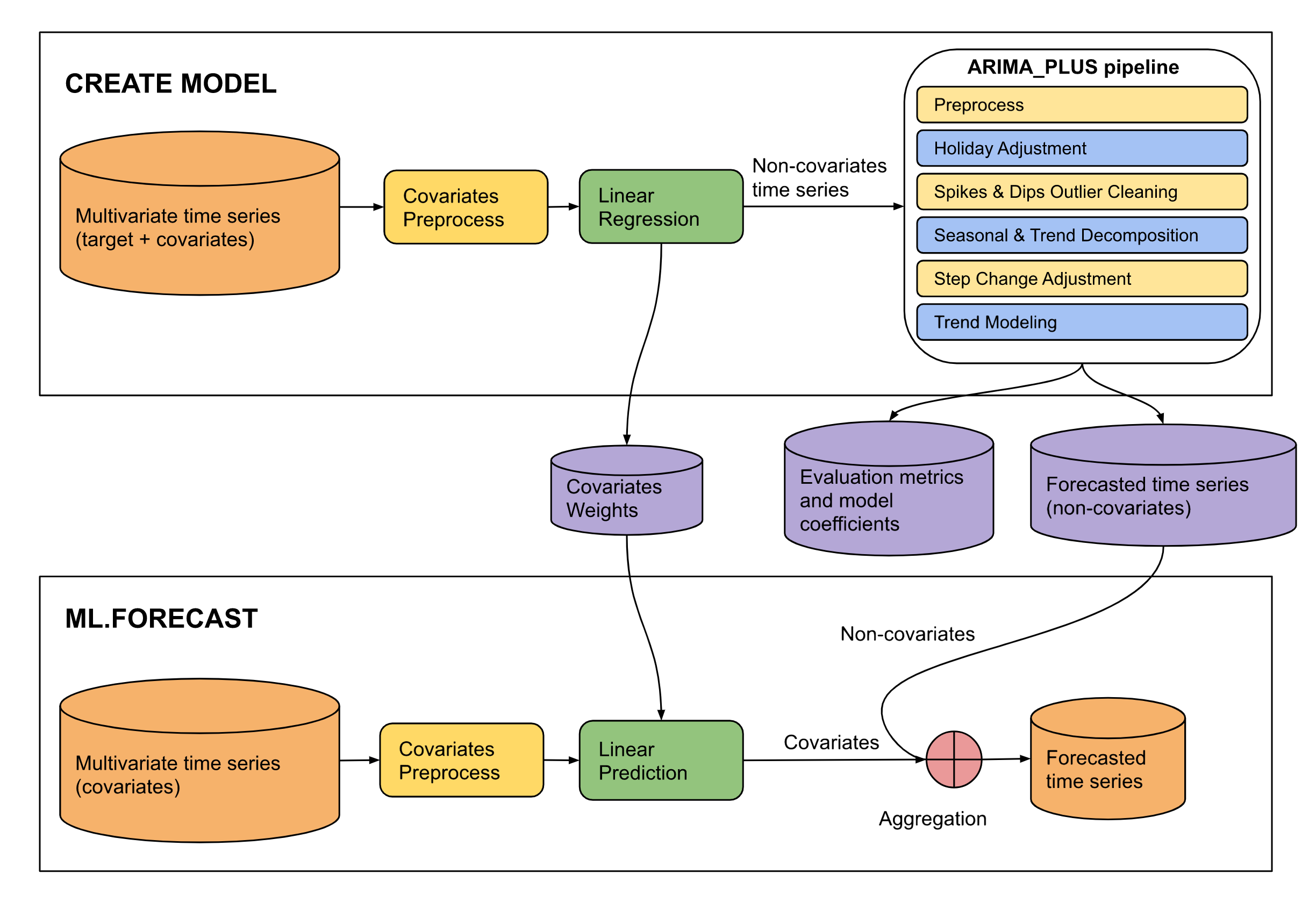}
\caption{The multivariate time series modeling pipeline, including model training and prediction. The ``CREATE MODEL" and ML functions in the diagram are discussed in Section \ref{interface}. The "ARIMA\_PLUS" pipeline refers to the univariate time series modeling pipeline, as illustrated in Figure \ref{fig-univariate}.
}\label{fig-multivariate}
\end{figure}

A simplified diagram is shown in Figure \ref{fig-multivariate}. The workflow is divided into two phases. The training stage (CREATE MODEL) first preprocesses the multivariate time series data. It then uses linear regression to separate the non-covariate series, which are subsequently processed through the univariate ARIMA\_PLUS pipeline. This stage outputs the linear weights, evaluation metrics, ARIMA model coefficients, and a non-covariate forecast. Following this, the forecasting stage (ML.FORECAST) takes new multivariate covariates, preprocesses them, applies linear prediction, and finally aggregates the result with the non-covariate forecast to yield the final time series forecasted values.

\subsection{Explaining forecast results}
In cases where the forecast deviates from actuals, or the visualization of the forecasting results seems to be problematic, we provide explanations for any forecasts that seem unusual or inaccurate. This is important for the stakeholders to have confidence in the forecasting results, as well as the practitioners to debug and further improve the results. 

Since our model architecture is based on time series decomposition, we can easily visualize the forecasts of each component such as trend, seasonality, holiday effect (both overall holiday effect and breakdown to the individual holiday level) and individual external regressor among others. This gives a clear picture of which component forecasts are reasonable while others may be problematic. 

In more detail, the historical time series is decomposed into the following components:

\begin{eqnarray}
history\_time\_series &=& trend + holiday\_e\textit{ff}ect \nonumber \\
& & + spikes\_and\_dips + step\_changes + residual \nonumber \\ & & + \sum _{m=\{daily,weekly,monthly,quarterly,yearly\}}seasonality_m \nonumber \\ 
& & + \sum _{i=1:N}attribution\_\textit{f}eature_i \nonumber
\end{eqnarray}

The forecasted data is decomposed into the following series. Compared to historical data, the forecasted data doesn’t have components including spikes and dips, step changes, and residuals.

\begin{eqnarray}
forecasted\_time\_series &=& trend + holiday\_e\textit{ff}ect \nonumber \\ & & + \sum _{m=\{daily,weekly,monthly,quarterly,yearly\}}seasonality_m \nonumber \\ & & 
+ \sum _{i=1:N}attribution\_\textit{f}eature_i \nonumber
\end{eqnarray}

\subsection{Anomaly detection}
\label{anomaly-detection}
In Section \ref{spikes-and-dips}, we discussed the approach to detect spike-and-dips anomalous points and removed them to achieve better forecasting results. In this section, we present a novel way to detect arbitrary anomalies in the time series. With this, we can achieve a unified framework to train a model that can do both forecasting and anomaly detection simultaneously. 

Note that ARIMA not only provides the expected values and upper/lower bounds in the future as part of the forecasting results, but also gives the expected values and upper/lower bounds for the trend time series, which can be used for historical data anomaly detection. The expected data value in each timestamp is computed as follows: $expected\_data = trend + holiday\_e\textit{ff}ect + seasonality + step\_changes$.

The residuals at each timestamp for historical data are assumed to follow the normal distribution $\mathcal{N}(0, \sigma)$, with zero mean and the standard error $\sigma$ obtained from the ARIMA fitting process. The anomaly probability identifies how far this point is away from the $expected\_data$ computed above. It is the probability that a random variable $X \sim \mathcal{N}(0, \sigma)$ is smaller than the difference between the expected value and the actual value, i.e.:
\[
P_{anomaly} =|1 - 2 \Phi(\frac{actual\_data - expected\_data}{\sigma})|
\]
where $\Phi$ is the cumulative distribution function of the  normal distribution. From the above equation, given the anomaly probability threshold $P_{th}$, the prediction interval size is computed as follows.
\[
S_{interval} = \Phi^{-1}(1 - \frac{P_{th}}{2}) \times \sigma
\]
The prediction lower bound $l$ and upper bound $u$ are then computed as: $[l,u] = [expected\_data - S_{interval}, expected\_data + S_{interval}]$. An actual time series data is an anomaly if it is not within the range of upper bound and lower bound. Equivalently, an actual time series data is an anomaly if the anomaly probability is higher than the anomaly probability threshold.

\subsection{Practical Forecasting Needs}
\subsubsection{Forecast within Limits}
In many use cases, the time series forecast results will fall out of a range that makes the forecasting results meaningless. For example, forecasting product demand to be under 0 is undesired. To solve this problem, we apply the algorithm in [9] to ensure the forecast is within the limits $[a, b]$. Note that $a$ could be negative infinity or $b$ could be infinity, making the range only one-sidedly bounded. To handle this, we extend the log transformation algorithm in [9]. When both $[a, b]$ are provided, then we can transform the data using a scaled logit transform which maps $(a , b)$ to the whole real line: $y = log\frac{x-a}{b-x}$ , where $x$ is on the original scale and $y$ is the transformed data. We forecast $y$ to get $y_{forecast}$ after transformation via the forecasting model as mentioned above. To reverse the transformation, we will use $x = \frac{(b - a)e^y}{1 + e^y} + a$.
For the forecasting intervals, it needs to be calculated as follows: the prediction interval on the transformed time series is $[y - c\times \sigma, y + c\times \sigma]$, where $c$ is the $z_{score}$ and $\sigma$ is the standard deviation. Mapping back to the original scale $x$, it would be 
$[\frac{a(b - x)e^{c\times \sigma} + b(x - a)}{(b - x) e^{c\times \sigma} + (x - a)}, \frac{a(b- x) + b(x - a)e^{c\times \sigma}}{(b- x)+ (x - a)e^{c\times \sigma}}]$. When only the lower bound a is provided, transformation $y = log(x - a)$ and the reverse transformation $x = e^y + a$ are used. Similarly, we can derive the prediction interval as 
$[a + e^{log(x - a) - c\times \sigma}, a +  e^{log(x - a) + c\times \sigma}]$
When only the upper bound b is provided, transformation $y = log(b - x)$ and the reserve transformation $x =b - e^y$ are used. Similarly, we can derive the prediction interval as 
$[b - e^{log(b - x) + c\times \sigma}, b - e^{log(b - x) - c\times \sigma}]$.

\subsubsection{Fast Trend Modeling}
The sequential nature of time series data necessitates a sequential processing approach during model training, where each data point is processed in chronological order. Consequently, model training time exhibits a linear relationship with the length of the time series, leading to increased computational costs for longer time series. Utilizing a subset of the data can therefore yield significant speed improvements.

While random downsampling is a common technique for reducing data size in independent and identically distributed data, it is unsuitable for time series due to their inherent temporal dependencies. Moreover, the contribution of data points to the modeling process is not uniform; more recent observations generally hold greater relevance. Consequently, an effective strategy for reducing input data size involves retaining only the most recent N data points.

Our contribution lies in the judicious use of a partial training dataset for specific time series component modeling. In detail, we restrict the subset of the input time series to trend modeling only, while the full data span is used for all other non-trend components. This methodology is justified by the following considerations:
\begin{enumerate}[label=(\arabic*)]
    \item Seasonality detection: Identifying seasonal patterns requires a sufficient time span. For instance, detecting yearly seasonality in daily data necessitates at least one year of observations. Using a subset may preclude the detection of such patterns, potentially compromising forecast accuracy.
    \item Computational cost: Modeling non-trend components is computationally much less demanding compared to ARIMA trend modeling. Therefore, the potential reduction in training time achieved by using a subset is often negligible.
\end{enumerate}



\section{System and Infrastructure}
\label{system-and-infra}
\subsection{Interface}
\label{interface}
\subsubsection{Model creation}
We provide a SQL interface to the model training, inference, evaluation and inspection. For training, it’s exposed by a CREATE MODEL query, with the full syntax provided below. For the explanation of different options in the query, refer to \cite{CreateModel}. The query can accept two input tables for creating the model, one is the input time series which is annotated as “training\_data” and the other is the customized holiday information which is annotated as ``custom\_holiday”. Specifying the model type as ARIMA\_PLUS indicates this is a univariate model where covariate features are disallowed in the training\_data; while ARIMA\_PLUS\_XREG specifies a multivariate model where covariate features are allowed. “TIME\_SERIES\_TIMESTAMP\_COL” and “TIME\_SERIES\_DATA\_COL” define a time series. The time series ID columns identify each individual time series, which means the input data can contain many time series. For each time series, an independent model will be fit. Other model options such as “auto\_arima” or “holiday\_region” are configuration parameters for each component in the training pipeline. Although the algorithms or models in the training pipeline have a lot of configuration parameters, they all have good default values which have been validated against a broad spectrum of time series data.

\begin{verbatim}
{CREATE MODEL | CREATE OR REPLACE MODEL}
model_name
OPTIONS( model_option_list )
AS { query_statement |
  (
    training_data AS ( query_statement ),
    custom_holiday AS ( holiday_statement )
  )
}

Model_option_list :
MODEL_TYPE = `ARIMA_PLUS | ARIMA_PLUS_XREG'
    [, TIME_SERIES_TIMESTAMP_COL = string_value ]
    [, TIME_SERIES_DATA_COL = string_value ]
    [, TIME_SERIES_ID_COL = string_value | string_array ]
    [, HORIZON = string_value ]
    [, AUTO_ARIMA = { TRUE | FALSE } ]
    [, AUTO_ARIMA_MAX_ORDER = int64_value ]
    [, AUTO_ARIMA_MIN_ORDER = int64_value ]
    [, NON_SEASONAL_ORDER = (int64_value, int64_value, int64_value) ]
    [, DATA_FREQUENCY = { `AUTO_FREQUENCY' | `PER_MINUTE' | `HOURLY' |
    `DAILY' | `WEEKLY' | `MONTHLY' | `QUARTERLY' | `YEARLY' } ]
    [, INCLUDE_DRIFT = { TRUE | FALSE } ]
    [, HOLIDAY_REGION = string_value | string_array ]
    [, CLEAN_SPIKES_AND_DIPS = { TRUE | FALSE } ]
    [, ADJUST_STEP_CHANGES = { TRUE | FALSE } ]
    [, TIME_SERIES_LENGTH_FRACTION = float64_value ]
    [, MIN_TIME_SERIES_LENGTH = int64_value ]
    [, MAX_TIME_SERIES_LENGTH = float64_value ]
    [, TREND_SMOOTHING_WINDOW_SIZE = float64_value ]
    [, DECOMPOSE_TIME_SERIES = { TRUE | FALSE } ]
    [, FORECAST_LIMIT_LOWER_BOUND = float64_value ]
    [, FORECAST_LIMIT_UPPER_BOUND = float64_value ]
    [, SEASONALITIES = string_array ]
    [, HIERARCHICAL_TIME_SERIES_COLS = { string_array } ]
    [, L2_REG = float64_value ]
\end{verbatim}

We show a simple example with most of the options set to the default values.
\begin{verbatim}
CREATE OR REPLACE MODEL `bqml_tutorial.liquor_forecast_by_product'
OPTIONS(
  MODEL_TYPE = `ARIMA_PLUS',
  TIME_SERIES_TIMESTAMP_COL = `date', 
  TIME_SERIES_DATA_COL = `total_bottles_sold',
  TIME_SERIES_ID_COL = [`city', `item_description'],
  HOLIDAY_REGION = `US'
) AS
SELECT 
  city, 
  item_description,
  date,
  SUM(bottles_sold) as total_bottles_sold
FROM    
  `bigquery-public-data.iowa_liquor_sales.sales'
GROUP BY city, item_description, date
\end{verbatim}

Running the above query will create a model entity in BigQuery. The model is named ``liquor\_forecast\_by\_product”, within the “bqml\_tutorial” dataset, which is a BigQuery concept for organizing data. 

\subsubsection{Forecasting and evaluation}
Once the model is created, there are a number of Table Value Functions (TVFs) that can be used to operate on the model, including model inference functions such as ML.FORECAST, ML.EXPLAIN\_FORECAST, ML.DETECT\_ANOMALIES, and model evaluation or inspection functions ML.EVALUATE and ML.ARIMA\_COEFFICIENTS.

Taking ML.FORECAST as an example, its syntax is shown below. For the ARIMA\_PLUS\_XREG multivariate model, new input data is required for the prediction of the linear model. For details, refer to \cite{MLForecast}.

\begin{verbatim}
# For ARIMA_PLUS:
ML.FORECAST(
  MODEL `project_id.dataset.model'
    STRUCT(
      [, 30 AS horizon]
      [, 0.95 AS confidence_level]))

# For ARIMA_PLUS_XREG:
ML.FORECAST(
  MODEL `project_id.dataset.model'
    { TABLE `project_id.dataset.table' | (query_statement) }
    STRUCT(
      [, 30 AS horizon]
      [, 0.95 AS confidence_level]))
\end{verbatim}

\subsection{Implementation}
The algorithm part of ARIMA\_PLUS, including ARIMA model training, STL, and ridge regression for covariates modeling, is entirely implemented in C\texttt{++}. The forecasting and evaluation functionalities, implemented as TVFs, are implemented using SQL query plans, therefore are seamlessly integrated into Dremel query engine \cite{MGL+10}. Furthermore, these components are orchestrated in a managed, serverless and distributed cloud environment, thereby ensuring fault tolerance and facilitating robust failure recovery.

\subsection{Handling multiple time series via resource auto-scaling}
\begin{figure}[H]
\centering
\includegraphics[width=0.95\textwidth]{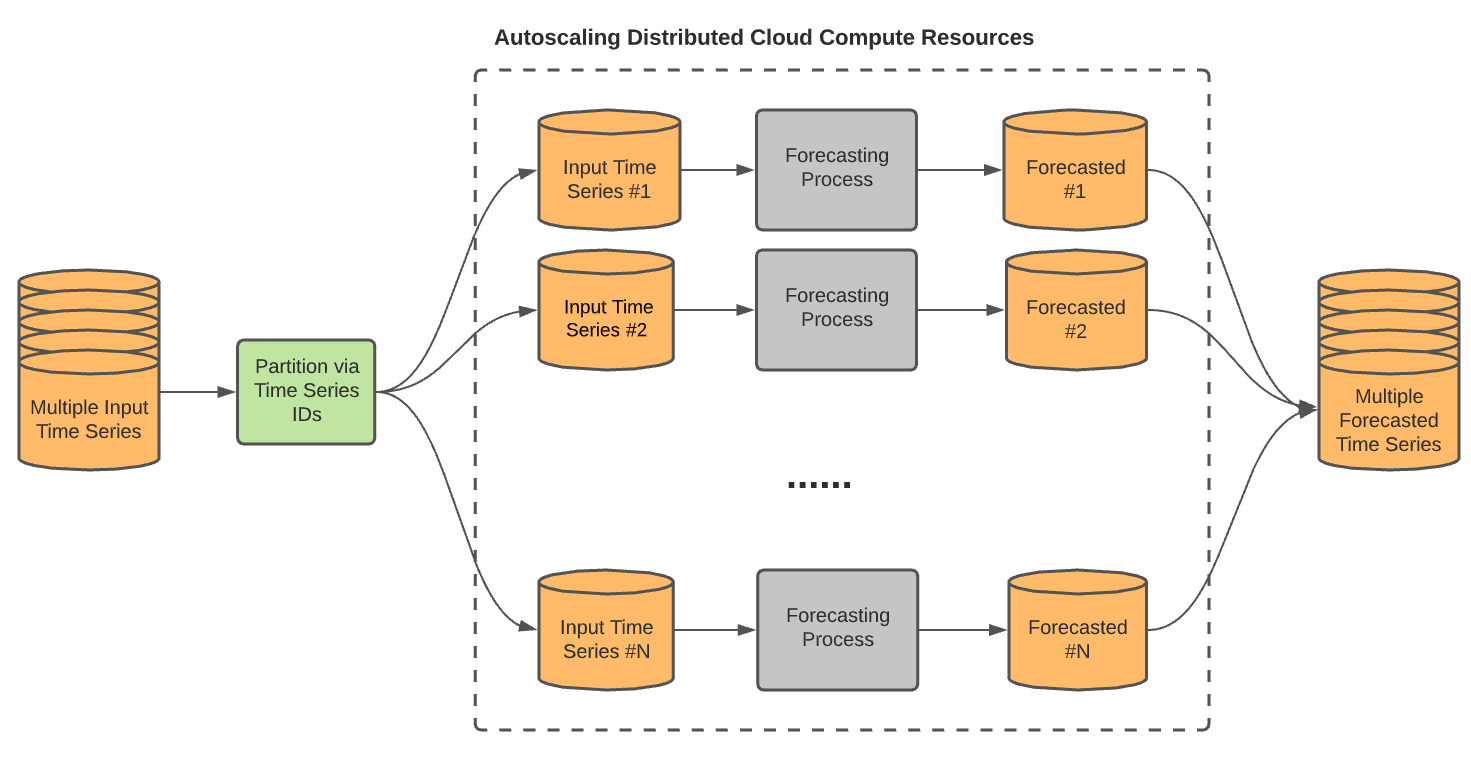}
\caption{For multiple time series, the system first identifies them using the ID columns. It then executes a separate training and prediction pipeline for each series, distributing the workload and auto-scaling compute and storage resources as needed.
}\label{fig-multiple-series}
\end{figure}

In real-world applications, it's common to have many (e.g., hundreds of millions of) time series to forecast, one for each SKU. It’s cumbersome to create hundreds of millions of time series models and manage them independently. To address this, we innovated a solution that automatically identifies each time series (univariate or multivariate) via one or more ID columns. This enables the parallel training and inference of every time series within a distributed environment, leveraging auto-scaling for cloud compute and storage resources. This is illustrated in Figure \ref{fig-multiple-series}. 

The generated millions of model artifacts are all stored in one model entity for easy management, which is discussed in the next section. 

\subsection{Model format and storage}
As mentioned earlier, running the CREATE MODEL query will create a model. The model file is a BigQuery table composed of rows and columns. 

The saved model must be self-contained to include all components required for the forecast or evaluation functions. We describe the model for ARIMA\_PLUS\_XREG, whose content is a super set of the ARIMA\_PLUS univariate model. The table should have the following parts for each time series:
(1) the historical univariate time series and its forecasted time series, (2) its decomposed time series components for history univariate data, (3) the ridge regression model weights for the covariate modeling, (4) a struct that holds the information about the ARIMA model fitting process, such as AIC, log likelihood, and variance.

When dealing with multiple time series, uniquely identified by one or more ID columns, the model table structure is expanded by appending new rows. All information corresponding to an individual time series model is sequentially stored within the rows that share a common time series ID. Consequently, a single BigQuery table can efficiently store a vast number of ARIMA\_PLUS models, each corresponding to an individual time series, which greatly simplifies overall model management.

\section{Experiments}
\label{experiments}
\subsection{Univariate forecasting accuracy}
\subsubsection{Full benchmark on Monash datasets}
For the quality benchmark we use all the 42 datasets from the well-known Monash forecasting repository. Some of the dataset information, such as time series frequency and the size of the dataset, are listed in Table \ref{monash-benchmark-all-metrics}. Since the time series in all the datasets are univariate time series, we use ARIMA\_PLUS univariate model for benchmarking. When training the ARIMA\_PLUS model, we use its default settings, including (1) auto.ARIMA being enabled with $auto\_arima\_max\_order = 2$; (2) only at most 1024 points for ARIMA trend modeling, and all points are used to model the non-trend components such as seasonality. The effect of some of above the hyperparameters on forecasting accuracy is evaluated separately in Section \ref{evaluate-hyperparameters}.

For evaluation, we compare the forecasted values against the actual values with predefined horizon in the Monash forecasting repository (see the horizon column in Table \ref{monash-benchmark-all-metrics}). We calculate the following error metrics: (1) mean absolute error (MAE), (2) root mean squared error (RMSE), (3) mean absolute percentage error (MAPE), (4) symmetric mean absolute percentage error (SMAPE), (5) mean absolute scaled error (MASE). The results are shown in Table \ref{monash-benchmark-all-metrics}, and can be reproduced by using this notebook which is published in Github \cite{monash-benchmark-notebook}. 

\begin{table}[H]
\centering
\resizebox{\textwidth}{!}{
\begin{tabular}{|c|c|c|c|c|c|c|c|c| }
  \hline
  Dataset & Frequency & Number of time series &
Horizon &
MAE & 
RMSE & 
MAPE & 
sMAPE &
MASE \\
\hline
Aus. Electricity demand & 
hourly & 
5 &
336 &
358.639 & 
417.744 & 
15.181 & 
12.908 & 
1.785 \\
\hline
Bitcoin & 
daily & 
18 &
30 &
9.404E+17 & 
1.129e+18 & 
9741.23 & 
32.018 &
5.798 \\
\hline
Carparts & 
monthly & 
2674 &
12 &
0.447 & 
0.755 & 
88.290 & 
188.478 &
0.790 \\
\hline
CIF 2016 & 
monthly & 
72 &
12 &
792193.87 & 
933910.47 & 
14.811 & 
16.453 &
0.957 \\
\hline
COVID & 
daily & 
266 &
30 &
159.827 & 
183.607 & 
14.754 & 
23.237 &
8.043 \\
\hline
Dominick & 
weekly & 
115704 &
8 &
5.340 & 
6.624 & 
67.285 & 
125.555 &
0.533 \\
\hline
Electricity Hourly & 
hourly & 
321 &
48 &
208.379 & 
297.452 & 
29.213 & 
16.040 &
1.358 \\
\hline
Electricity Weekly & 
weekly & 
321 &
8 &
21840.84 & 
25323.58 & 
9.886 & 
8.872 &
0.732 \\
\hline
FRED-MD & 
monthly & 
107 &
12 &
2803.46 & 
3135.90 & 
31.676 & 
11.01 &
0.547 \\
\hline
Hospital & 
monthly & 
767 &
12 &
18.614 & 
22.580 & 
20.186 & 
18.085 &
0.790 \\
\hline
Kaggle Weekly & 
weekly & 
145063 &
8 &
2299.59 & 
2980.36 & 
99.802 & 
41.752 & 
0.616 \\
\hline
KDD & 
hourly & 
270 &
48 &
32.682 & 
40.41 & 
136.197 & 
61.272 &
1.107 \\
\hline
M1 Monthly & 
monthly & 
617 &
18 &
1954.321 & 
2313.73 & 
18.798 & 
16.623 &
1.118 \\
\hline
M1 Quarterly & 
quarterly & 
203 &
8 &
2003.64 & 
2366.36 & 
16.751 & 
16.322 &
1.622 \\
\hline
M1 Yearly & 
yearly & 
181 &
6 &
121751.68 & 
139974.80 & 
17.080 & 
16.778 &
3.429 \\
\hline
M3 Monthly & 
monthly & 
1428 &
18 &
655.247 & 
802.14 & 
20.38 & 
15.563 &
0.882 \\
\hline
M3 Other & 
quarterly & 
174 &
8 &
206.515 & 
239.66 & 
4.993 & 
4.646 &
0.764 \\
\hline
M3 Quarterly & 
quarterly & 
756 &
8 &
518.42 & 
603.08 & 
12.669 & 
9.997 &
1.159 \\
\hline
M3 Yearly & 
yearly & 
645 &
6 &
1191.60 & 
1365.57 & 
22.748 & 
17.745 &
3.009 \\
\hline
M4 Hourly & 
hourly & 
414 &
48 &
298.309 & 
369.253 & 
15.96 & 
15.136 &
1.074 \\
\hline
M4 Daily & 
daily & 
4227 &
14 &
232.759 & 
263.122 & 
4.979 & 
4.299 &
1.515 \\
\hline
M4 Weekly & 
weekly & 
359 &
13 &
294.203 & 
352.743 & 
7.594 & 
7.417 &
0.491 \\
\hline
M4 Monthly & 
monthly & 
48000 &
18 &
565.718 & 
691.23 & 
15.771 & 
13.587 &
0.939 \\
\hline
M4 Quarterly & 
quarterly & 
24000 &
8 &
595.258 & 
696.787 & 
12.616 & 
10.980 &
1.214 \\
\hline
M4 Yearly & 
yearly & 
22974 &
6 &
931.76 & 
1059.08 & 
18.425 & 
15.493 &
3.431 \\
\hline
NN5 Daily & 
daily & 
111 &
56 &
4.053 & 
5.699 & 
64.781 & 
24.801 &
0.943 \\
\hline
NN5 Weekly & 
weekly & 
111 &
8 &
14.308 & 
17.729 & 
11.384 & 
11.050 &
0.823 \\
\hline
Pedestrians & 
hourly & 
66 &
24 &
59.687 & 
82.482 & 
62.327 & 
73.195 &
0.340 \\
\hline
Rideshare & 
hourly & 
2304 &
48 &
1.126 & 
1.486 & 
23.21 & 
26.85 &
0.712 \\
\hline
Saugeen & 
daily & 
1 &
30 &
26.401 & 
48.705 & 
34.977 & 
49.896 &
1.751 \\
\hline
Solar 10 Mins & 
hourly & 
137 &
60 &
0.517 & 
1.377 & 
100 & 
200 &
0.229 \\
\hline
Solar Weekly & 
weekly & 
137 &
5 &
1315.79 & 
1467.94 & 
32.197 & 
26.317 &
1.324 \\
\hline
Sunspot & 
daily & 
1 &
30 &
0.0667 & 
0.365 & 
100 & 
200 &
0.00173 \\
\hline
Temp. Rain & 
daily & 
32072 &
30 &
6.863 & 
9.432 & 
19456.80 & 
154.959 &
0.884 \\
\hline
Tourism Monthly & 
monthly & 
366 &
24 &
2177.171 & 
2893.046 & 
22.536 & 
20.899 &
1.644 \\
\hline
Tourism Quarterly & 
quarterly & 
427 &
8 &
7392.460 & 
8765.128 & 
17.012 & 
16.672 &
1.662 \\
\hline
Tourism Yearly & 
yearly & 
518 &
4 &
74587.897 & 
80921.066 & 
34.230 & 
33.617 &
3.364 \\
\hline
Traffic Hourly & 
hourly & 
862 &
168 &
0.0150 & 
0.0242 & 
50.587 & 
51.778 &
0.964 \\
\hline
Traffic Weekly & 
weekly & 
862 &
8 &
1.152 & 
1.544 & 
24.881 & 
12.656 &
1.125 \\
\hline
US Births & 
daily & 
1 &
30 &
425.621 & 
538.360 & 
4.323 & 
4.272 &
1.550 \\
\hline
Vehicle Trips & 
daily & 
329 &
30 &
21.881 & 
26.664 & 
96.488 & 
32.953 &
1.989 \\
\hline
Weather & 
daily & 
3010 &
30 &
1.995 & 
2.728 & 
70.475 & 
61.610 &
0.610 \\
  \hline
\end{tabular}
}
\caption{Forecasting accuracy of the ARIMA\_PLUS model in various evaluation metrics on the 42 datasets of the Monash forecasting repository. For each dataset, we list its data frequency, the number of time series, forecasting horizon, and the error metrics.}
\label{monash-benchmark-all-metrics}
\end{table}

We compare the forecasting performance of ARIMA\_PLUS against a variety of time series models using the MASE metric. Although MAPE and RMSE are more well-known, MASE is the preferred metric used in the M4 competition \cite{MSA20}. Among other benefits shown in \cite{HK06}, MASE is scale-independent, making it ideal for comparing the accuracy of forecast models across different time series with varying scales or units. To facilitate this comparison, we leverage the metrics of other models reported on the Monash forecasting repository. The MASE result, per-dataset, are shown in Table \ref{monash-benchmark-mase}. 
\clearpage
\begin{landscape}
\begin{table}[H]
\resizebox{\linewidth}{!}{
\begin{tabular}{|l|c|c|c|c|c|c|c|c|c|c|c|c|c|c|c|c|c|c|c|c|c|c|c|c|c|}
  \hline
  Dataset & ARIMA\_PLUS & SES & Theta & TBATS & ETS & (DHR-)ARIMA & PR & CatBoost & FFNN & DeepAR & N-BEATS & WaveNet & Transformer & Prophet & DLinear & TTM (ZS) & iTransformer & N-BEATS & N-HITS & NLinear & PatchTSMixer & PatchTST & TiDE & TimeMixer & TSMixer \\
  \hline
Aus. Elecdemand & 1.785 & 1.857 & 1.867 & 1.174 & 5.663 & 2.574 & \textbf{0.780} & \textbf{0.705} & 1.222 & 1.591 & \textbf{1.014} & 1.102 & 1.113 & 1.414 & 1.158 & \na & 1.346 & 1.094 & 1.143 & 1.142 & 1.068 & 1.182 & 1.217 & 1.121 & 1.302 \\
\hline
Bitcoin & 5.798 & 4.327 & 4.344 & 4.611 & \textbf{2.718} & 4.030 & \textbf{2.664} & \textbf{2.888} & 6.006 & 6.394 & 7.254 & 5.315 & 8.462 & 11.089 & 6.958 & 4.864 & \na & 6.408 & 6.066 & 5.746 & \na & \na & \na & \na & 6.441 \\
\hline
Carparts & 0.790 & 0.897 & 0.914 & 0.998 & 0.925 & 0.926 & 0.755 & 0.853 & \textbf{0.747} & \textbf{0.747} & 2.836 & 0.754 & \textbf{0.746} & 0.876 & 0.747 & 0.914 & 0.839 & 0.753 & 0.748 & 1.052 & 0.802 & 0.890 & 0.747 & 0.757 & 0.784 \\
\hline
CIF 2016 & 0.957 & 1.291 & 0.997 & \textbf{0.861} & \textbf{0.841} & \textbf{0.929} & 1.019 & 1.175 & 1.053 & 1.159 & 0.971 & 1.800 & 1.173 & 1.029 & 3.717 & 1.216 & 1.316 & 2.261 & 1.622 & 1.333 & 1.650 & 1.230 & 2.352 & 1.360 & 1.302 \\
\hline
COVID & 8.043 & 7.776 & 7.793 & 5.719 & 5.326 & 6.117 & 8.731 & 8.241 & 5.459 & 6.895 & 5.858 & 7.835 & 8.941 & 12.770 & 5.974 & 7.603 & \textbf{5.197} & 7.911 & 7.362 & \textbf{5.233} & 5.300 & \textbf{5.031} & 9.069 & 5.367 & 5.542 \\
\hline
Dominick & 0.533 & 0.582 & 0.610 & 0.722 & 0.595 & 0.796 & 0.980 & 1.038 & 0.614 & 0.540 & 0.952 & 0.531 & 0.531 & 0.827 & 0.546 & \textbf{0.485} & 0.525 & \textbf{0.511} & \textbf{0.507} & 0.544 & 0.537 & 0.553 & 0.550 & 0.517 & 0.517 \\
\hline
Electricity Hourly & \textbf{1.358} & 4.544 & 4.545 & 3.690 & 6.501 & 4.602 & 2.912 & 2.262 & 3.200 & 2.516 & 1.968 & \textbf{1.606} & 2.522 & 2.050 & 2.012 & \na & 3.291 & 2.005 & 2.073 & \textbf{1.949} & 2.841 & 2.571 & 2.127 & 1.930 & 2.074 \\
\hline
Electricity Weekly & \textbf{0.732} & 1.536 & 1.476 & 0.792 & 1.526 & 0.878 & 0.916 & 0.815 & \textbf{0.769} & 1.005 & 0.800 & 1.250 & 1.770 & 0.924 & 0.827 & 0.887 & 0.972 & 0.884 & 0.986 & \textbf{0.797} & 0.827 & 0.845 & 0.811 & 0.991 & 0.792 \\
\hline
FRED-MD & 0.547 & 0.617 & 0.698 & 0.502 & \textbf{0.468} & 0.533 & 8.827 & 0.947 & 0.601 & 0.640 & 0.604 & 0.806 & 1.823 & 1.843 & 0.735 & 0.689 & 0.749 & 0.666 & 0.660 & 0.848 & 0.591 & 0.635 & 0.901 & 0.617 & \textbf{0.587} \\
\hline
Hospital & 0.790 & 0.813 & \textbf{0.761} & 0.768 & \textbf{0.765} & 0.787 & 0.782 & 0.798 & 0.840 & 0.769 & 0.791 & 0.779 & 1.031 & \textbf{0.673} & 0.808 & 0.791 & 0.866 & 0.786 & 0.802 & 0.809 & 0.801 & 0.811 & 0.850 & 0.811 & 0.818 \\
\hline
Kaggle Weekly & 0.616 & 0.698 & 0.694 & 0.622 & 0.770 & 0.815 & 1.021 & 1.928 & 0.689 & 0.758 & 0.667 & 0.628 & 0.888 & 1.196 & 0.715 & \textbf{0.544} & 0.620 & 0.613 & \textbf{0.582} & 0.663 & 0.612 & 0.687 & 0.720 & 0.615 & \textbf{0.595} \\
\hline
KDD & \textbf{1.107} & 1.645 & 1.646 & 1.394 & 1.787 & 1.982 & 1.265 & 1.233 & 1.228 & 1.699 & 1.600 & \textbf{1.185} & 1.696 & \textbf{1.186} & 1.280 & \na & 1.331 & 1.252 & 1.194 & 1.256 & 1.229 & 1.225 & 1.304 & 1.245 & 1.244 \\
\hline
M1 Monthly & 1.118 & 1.379 & \textbf{1.091} & 1.118 & \textbf{1.074} & 1.164 & 1.123 & 1.209 & 1.205 & 1.192 & \textbf{1.168} & 1.200 & 2.191 & 1.712 & 1.535 & 1.309 & 1.433 & 1.253 & 1.257 & 1.706 & 1.268 & 1.381 & 1.501 & 1.265 & 1.255 \\
\hline
M1 Quarterly & \textbf{1.622} & 1.929 & 1.702 & 1.694 & \textbf{1.658} & 1.787 & 1.892 & 2.031 & 1.862 & 1.833 & 1.788 & \textbf{1.700} & 2.772 & 2.136 & 3.107 & 1.864 & 1.945 & 2.543 & 2.369 & 3.721 & 1.891 & 2.064 & 2.777 & 1.866 & 1.892 \\
\hline
M1 Yearly & \textbf{3.429} & 4.938 & 4.191 & \textbf{3.499} & \textbf{3.771} & 4.479 & 4.588 & 4.427 & 4.355 & 4.603 & 4.384 & 4.666 & 5.519 & 5.633 & 5.724 & 4.656 & 4.052 & 5.525 & 4.840 & 8.475 & 4.170 & 4.331 & 6.258 & 4.051 & 4.114 \\
\hline
M3 Monthly & 0.882 & 1.091 & \textbf{0.864} & \textbf{0.861} & \textbf{0.865} & 0.873 & 1.010 & 1.065 & 1.011 & 1.167 & 0.934 & 1.008 & 1.454 & 1.375 & 1.079 & 1.024 & 1.051 & 0.916 & 0.906 & 0.989 & 1.005 & 1.080 & 1.066 & 0.947 & 1.000 \\
\hline
M3 Other & \textbf{0.764} & 3.089 & 2.271 & 1.848 & \textbf{1.814} & \textbf{1.831} & 2.655 & 3.178 & 2.615 & 2.975 & 2.390 & 2.127 & 2.781 & 4.694 & \na & \na & \na & \na & \na & \na & \na & \na & \na & \na & \na \\
\hline
M3 Quarterly & 1.159 & 1.417 & \textbf{1.117} & 1.256 & 1.170 & 1.240 & 1.248 & 1.441 & 1.329 & 1.310 & 1.182 & 1.290 & 2.452 & 1.672 & 1.254 & 1.338 & 1.262 & \textbf{1.134} & \textbf{1.136} & 1.209 & 1.208 & 1.261 & 1.522 & 1.139 & 1.162 \\
\hline
M3 Yearly & 3.009 & 3.167 & \textbf{2.774} & 3.127 & 2.860 & 3.417 & 3.223 & 3.788 & 3.399 & 3.508 & 2.961 & 3.014 & 3.003 & 4.152 & 2.993 & 3.004 & \textbf{2.956} & 3.024 & \textbf{2.824} & 3.020 & 2.962 & 3.104 & 3.272 & 2.963 & 2.961 \\
\hline
M4 Daily & 1.515 & 1.154 & 1.153 & 1.157 & 1.239 & 1.179 & 1.162 & 1.593 & \textbf{1.141} & 2.212 & 1.218 & 1.157 & 1.377 & 3.698 & \textbf{1.100} & \textbf{1.066} & 1.245 & 1.125 & 1.142 & 1.146 & 1.150 & 1.187 & 1.324 & 1.150 & 1.136 \\
\hline
M4 Hourly & \textbf{1.074} & 11.607 & 11.524 & 2.663 & 26.690 & 13.557 & \textbf{1.662} & 1.771 & 2.862 & 2.145 & 2.247 & \textbf{1.680} & 8.840 & 1.776 & 3.246 & 2.340 & 6.357 & 2.756 & 3.781 & 1.973 & 2.305 & 2.847 & 3.081 & 7.246 & 3.235 \\
\hline
M4 Monthly & \textbf{0.939} & 1.150 & \textbf{0.970} & 1.053 & \textbf{0.948} & 0.962 & 1.080 & 1.093 & 1.151 & 1.163 & 1.026 & 1.160 & 2.125 & 1.367 & 1.007 & 1.082 & 1.172 & 0.970 & 0.987 & 1.012 & 1.027 & 1.106 & 1.122 & 1.013 & 1.013 \\
\hline
M4 Quarterly & 1.214 & 1.417 & 1.231 & 1.186 & \textbf{1.161} & 1.228 & 1.316 & 1.338 & 1.420 & 1.274 & 1.239 & 1.242 & 1.520 & 1.758 & 1.260 & 1.389 & 1.439 & \textbf{1.209} & \textbf{1.226} & 1.330 & 1.325 & 1.410 & 1.469 & 1.311 & 1.306 \\
\hline
M4 Weekly & 0.491 & 0.587 & 0.546 & 0.504 & 0.575 & 0.550 & 0.481 & 0.615 & 0.545 & 0.586 & \textbf{0.453} & 0.587 & 0.695 & 1.049 & 0.497 & 0.631 & 0.744 & 0.484 & \textbf{0.473} & 0.488 & 0.531 & 0.537 & 0.560 & \textbf{0.478} & 0.503 \\
\hline
M4 yearly & 3.431 & \na & \na & \na & \na & \na & \na & \na & \na & \na & \na & \na & \na & \na & \textbf{3.343} & 3.727 & 3.433 & 3.430 & \textbf{3.344} & 3.800 & 3.433 & 4.124 & 3.698 & \textbf{3.427} & 3.400 \\
\hline
NN5 Daily & 0.943 & 1.521 & 0.885 & \textbf{0.858} & \textbf{0.865} & 1.013 & 1.263 & 0.973 & 0.941 & 0.919 & 1.134 & 0.916 & 0.958 & \textbf{0.883} & 0.958 & 1.487 & 0.949 & 0.901 & 0.908 & 0.959 & 0.965 & 1.016 & 0.962 & 0.934 & 0.942 \\
\hline
NN5 Weekly & \textbf{0.823} & 0.903 & 0.885 & 0.872 & 0.911 & 0.887 & 0.854 & 0.853 & 0.850 & 0.863 & \textbf{0.808} & 1.123 & 1.141 & 0.927 & 0.862 & 0.917 & 0.833 & 0.824 & 0.864 & 0.859 & \textbf{0.818} & 0.841 & 0.848 & 0.917 & 0.893 \\
\hline
Pedestrians & 0.340 & 0.957 & 0.958 & 1.297 & 1.190 & 3.947 & 0.256 & 0.262 & 0.267 & 0.272 & 0.380 & \textbf{0.247} & 0.274 & 2.034 & 0.258 & 0.304 & 0.396 & 0.245 & \textbf{0.241} & 0.252 & 0.263 & 0.261 & \textbf{0.245} & 0.261 & 0.254 \\
\hline
Rideshare & \textbf{0.712} & 3.014 & 3.641 & 3.067 & 4.040 & \textbf{1.530} & 3.019 & \textbf{2.908} & 4.198 & 4.029 & 3.877 & 3.009 & 4.040 & 4.666 & 4.114 & \na & 6.985 & 4.108 & 4.041 & 4.159 & 5.743 & 4.636 & 4.164 & 5.366 & 3.572 \\
\hline
Saugeen & 1.751 & 1.426 & 1.425 & 1.477 & 2.036 & 1.485 & 1.674 & \textbf{1.411} & 1.524 & 1.560 & 1.852 & 1.471 & 1.861 & 1.510 & 1.644 & 1.657 & 1.621 & 1.587 & 1.668 & 1.490 & \textbf{1.481} & 1.552 & 1.536 & 1.576 & 1.515 \\
\hline
Solar 10 Mins & \textbf{0.229} & 1.451 & 1.452 & 3.936 & 1.451 & \textbf{1.034} & 1.451 & 2.504 & 1.450 & 1.450 & 1.573 & \na & 1.451 & 1.821 & 1.620 & \na & \textbf{1.422} & 2.379 & 2.406 & 1.772 & 1.454 & 1.438 & 1.473 & 1.440 & 1.413 \\
\hline
Solar Weekly & 1.324 & 1.215 & 1.224 & 0.916 & 1.134 & 0.848 & 1.053 & 1.530 & 1.045 & \textbf{0.725} & 1.184 & 1.961 & \textbf{0.574} & 1.508 & 1.152 & 0.927 & 1.015 & 1.306 & 1.384 & 1.251 & 1.030 & 1.305 & 1.099 & 1.025 & 1.112 \\
\hline
Sunspot & \textbf{0.00173} & 0.128 & 0.128 & 0.067 & 0.128 & 0.067 & 0.099 & \textbf{0.059} & 0.207 & 0.020 & 0.375 & 0.004 & \textbf{0.003} & 0.852 & 0.113 & 0.123 & 0.078 & 0.192 & 0.166 & 0.089 & 0.063 & 0.098 & 0.174 & 0.065 & 0.079 \\
\hline
Temp. Rain & 0.884 & 1.347 & 1.368 & 1.227 & 1.401 & 1.174 & 0.876 & 1.028 & 0.847 & \textbf{0.785} & 1.300 & 0.786 & \textbf{0.687} & 1.150 & 1.214 & 0.893 & 0.837 & 0.791 & 0.795 & 1.374 & 0.854 & 1.159 & 1.201 & \textbf{0.784} & 0.820 \\
\hline
Tourism Monthly & 1.644 & 3.306 & 1.649 & 1.751 & 1.526 & 1.589 & 1.678 & 1.699 & 1.582 & \textbf{1.409} & 1.574 & 1.482 & 1.571 & 2.008 & 1.492 & 3.027 & 2.561 & \textbf{1.465} & \textbf{1.476} & 1.521 & 1.696 & 2.126 & 1.705 & 1.547 & 1.580 \\
\hline
Tourism Quarterly & 1.662 & 3.210 & 1.661 & 1.835 & 1.592 & 1.782 & 1.643 & 1.793 & 1.678 & 1.597 & \textbf{1.475} & 1.714 & 1.859 & 2.153 & \textbf{1.540} & 3.105 & 2.273 & 1.570 & \textbf{1.535} & 1.636 & 1.610 & 2.440 & 2.844 & 1.683 & 1.633 \\
\hline
Tourism Yearly & 3.364 & 3.253 & 3.015 & 3.685 & 3.395 & 3.775 & 3.516 & 3.553 & 3.401 & 3.205 & 2.977 & 3.624 & 3.552 & \textbf{2.590} & 6.863 & 3.078 & 2.757 & 109.908 & 3.195 & 6.454 & 2.836 & 3.362 & 5.358 & \textbf{2.710} & \textbf{2.733} \\
\hline
Traffic Hourly & 0.964 & 1.922 & 1.922 & 2.482 & 2.294 & 2.535 & 1.281 & 1.571 & 0.892 & \textbf{0.825} & 1.100 & 1.066 & \textbf{0.821} & 1.316 & 0.918 & \na & 0.888 & 0.873 & 0.850 & 0.913 & 1.075 & 0.946 & 0.924 & 0.842 & \textbf{0.816} \\
\hline
Traffic Weekly & 1.125 & 1.116 & 1.121 & 1.148 & 1.125 & 1.191 & 1.122 & 1.116 & 1.150 & 1.182 & \textbf{1.094} & 1.233 & 1.555 & \textbf{1.084} & 1.130 & 1.173 & 1.186 & 1.117 & \textbf{1.083} & 1.098 & 1.117 & 1.092 & 1.124 & 1.118 & 1.099 \\
\hline
US Births & 1.550 & 4.343 & 2.138 & \textbf{1.453} & 1.529 & 1.917 & 2.094 & 1.609 & 2.032 & 1.548 & 1.537 & 1.837 & 1.650 & 5.626 & 2.171 & 4.357 & 1.850 & \textbf{1.438} & \textbf{1.510} & 2.127 & 2.337 & 2.268 & 2.723 & 1.504 & 1.828 \\
\hline
Vehicle Trips & 1.989 & 1.224 & 1.244 & 1.860 & 1.305 & 1.282 & \textbf{1.212} & \textbf{1.176} & 1.843 & 1.929 & 2.143 & 1.851 & 2.532 & 2.428 & 1.954 & 1.984 & 2.018 & 1.747 & 1.728 & 1.934 & 1.963 & 2.002 & 2.015 & 1.890 & 1.912 \\
\hline
Weather & \textbf{0.610} & 0.677 & 0.749 & 0.689 & 0.702 & 0.746 & 3.046 & 0.762 & 0.638 & \textbf{0.631} & 0.717 & 0.721 & 0.650 & 0.880 & 0.640 & \textbf{0.571} & 0.650 & 0.625 & 0.621 & 0.641 & 0.650 & 0.640 & 0.648 & 0.638 & 0.656 \\
  \hline
\end{tabular}
}
\caption{Forecasting accuracy of the ARIMA\_PLUS model (2nd column) against a variety of time series models using the MASE metric, on the 42 datasets of the Monash time series repository. ``N/A" means that particular data isn't reported in the Monash website. We highlight the top three models with the lowest MASE values for each dataset in bold.
}\label{monash-benchmark-mase}
\end{table}
\end{landscape}

\begin{figure}[H]
\centering
\includegraphics[width=0.95\textwidth]{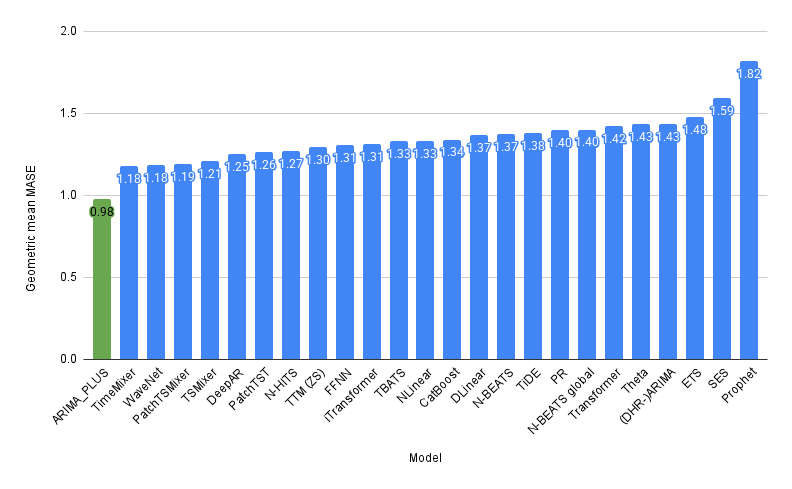}
\caption{Geometric mean MASE of ARIMA\_PLUS and other models on the Monash datasets. ARIMA\_PLUS achieves the lowest geometric mean MASE, indicating superior performance.
}\label{fig-mean-mase}
\end{figure}

Across the Monash datasets, we compute the geometric mean of MASE, which is considered more appropriate than arithmetic mean \cite{DKSZ23}. We plot the geometric mean MASE for all the time series models in Figure \ref{fig-mean-mase}. ARIMA\_PLUS outperforms all other models by a good margin; it is 17\% more accurate than the second-best model. Note that out  of  the  42  datasets,  the  “M4  yearly”  dataset  is  not  used  as  the  metric  is missing for most time series models from the Monash website.

\begin{figure}[H]
\centering
\includegraphics[width=0.95\textwidth]{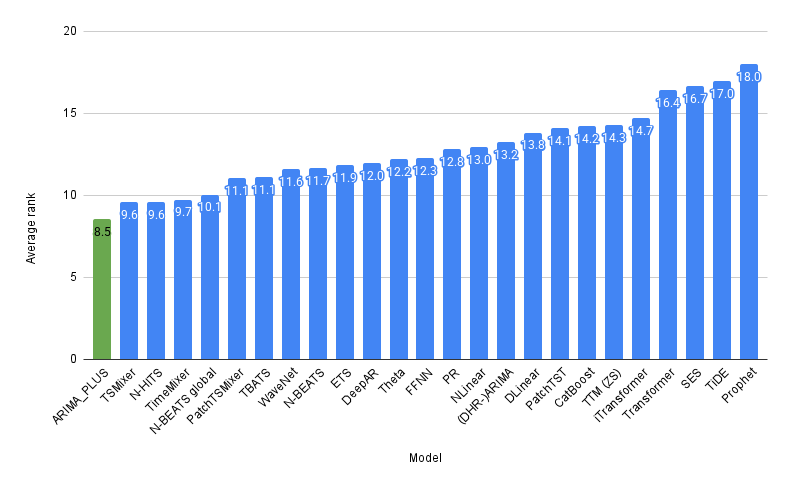}
\caption{Average rank of the MASE metric comparing ARIMA\_PLUS and other models across the Monash datasets. ARIMA\_PLUS achieves the highest rank among all the models. 
}\label{fig-average-rank}
\end{figure}

We also compute the average rank of MASE across the Monash datasets. Again, the  ``M4  yearly”  dataset  is  not  used  as  the  metric is missing for most time series models from the Monash website. To compute average rank, we first rank each model for every dataset based on the MASE score, and then perform an average over the rank scores for each model. The results are shown in Figure \ref{fig-average-rank}, where we can see that ARIMA\_PLUS ranks higher than all other models by a good margin: 8.5 vs 9.6 (2nd best). Beyond average rank, ARIMA\_PLUS secured the top 1 rank in 11 datasets and the top 5 rank in 19 datasets out of 41 total, outperforming all other models in both metrics.

\subsubsection{Evaluate hyperparameters in ARIMA trend modeling}
\label{evaluate-hyperparameters}
\begin{figure}[H]
\centering
\includegraphics[width=0.95\textwidth]{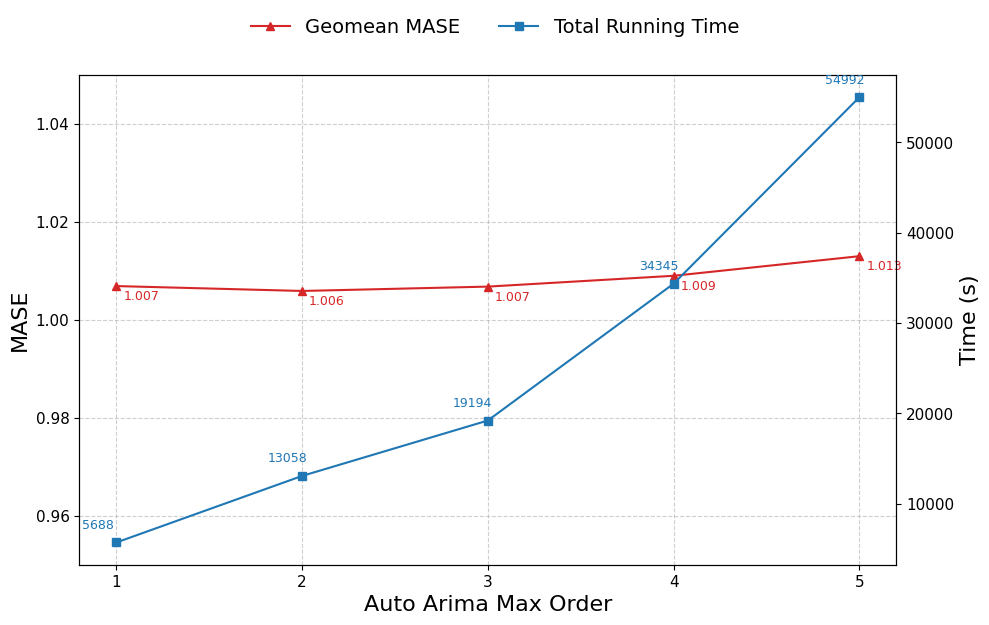}
\caption{Geometric mean $\text{MASE}$ of $\text{ARIMA\_PLUS}$ on the Monash datasets and the total model training time in seconds across all datasets. These results are shown with respect to different values of $auto\_arima\_max\_order$, which defines the search space for the $\text{auto.ARIMA}$ algorithm.
}\label{fig-auto-arima-max-order}
\end{figure}

We evaluate two hyperparameters' impact to the forecasting accuracy, the $auto\_arima\_max\_order$ (sum of ARIMA's $p$ and $q$) and the ``context window length" for ARIMA trend modeling. For $auto\_arima\_max\_order$, the results are shown in Figure \ref{fig-auto-arima-max-order}. We can see that increasing the search space of the auto.ARIMA algorithm doesn't much affect the overall forecasting accuracy, yet significantly increases the model fitting time. Paradoxically, expanding the auto.ARIMA search space results in a slightly higher MASE score and thus reduced forecasting accuracy. This is attributed to two factors: (1) the algorithm performs only a local optimization on the trend component, and (2) its selection criterion, AIC, is not directly optimized for forecasting metrics. This is the reason why we default $auto\_arima\_max\_order$ to 2. 

\begin{figure}[H]
\centering
\includegraphics[width=0.95\textwidth]{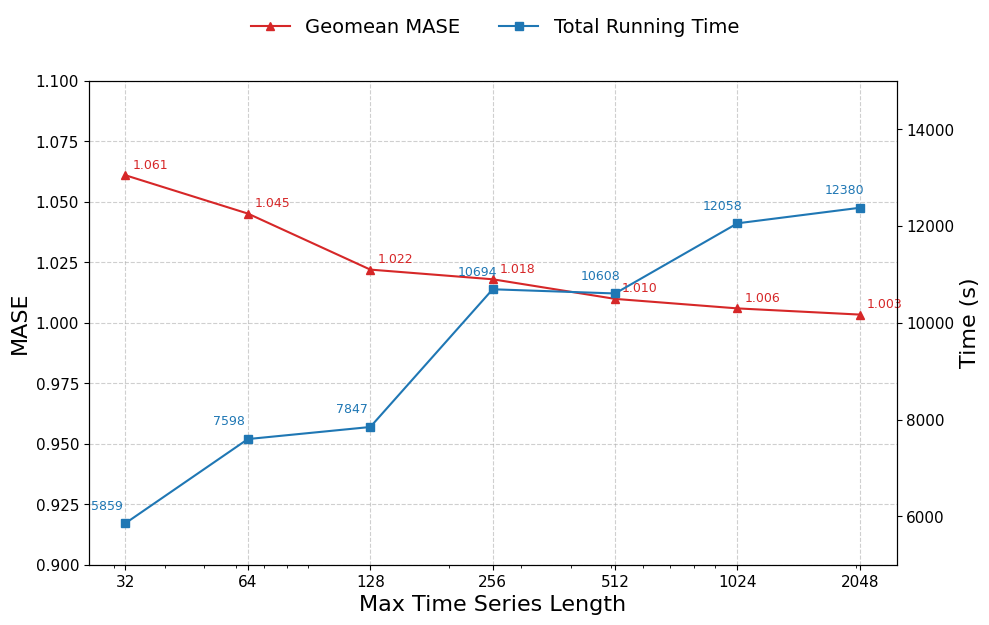}
\caption{geometric mean $\text{MASE}$ of $\text{ARIMA\_PLUS}$ on the Monash datasets and the total model training time in seconds across all datasets. These results are shown with respect to different values of $\text{MAX\_TIME\_SERIES\_LENGTH}$, which defines the context window used in $\text{ARIMA}$ trend modeling for each time series.
}\label{fig-context-window}
\end{figure}

The evaluation results of various context window lengths for ARIMA trend modeling are shown in Figure \ref{fig-context-window}. From the figure, we can see that the MASE score only improves by 1.86\% when the context window length is increased from 128 to 2048, but the time cost in training the models increases by around 58\%. This is the reason why we default its value to 1024.

\subsection{Multivariate forecasting accuracy}
For multivariate time series, in general utilizing the side features helps with improving the forecasting accuracy. We use an example to show case how the ARIMA\_PLUS\_XREG model outperforms ARIMA\_PLUS in forecasting accuracy for a multivariate time series. In this example, we use BigQuery public dataset ``\url{bigquery-public-data.epa\_historical\_air\_quality}". 

\begin{figure}[H]
\centering
\includegraphics[width=0.95\textwidth]{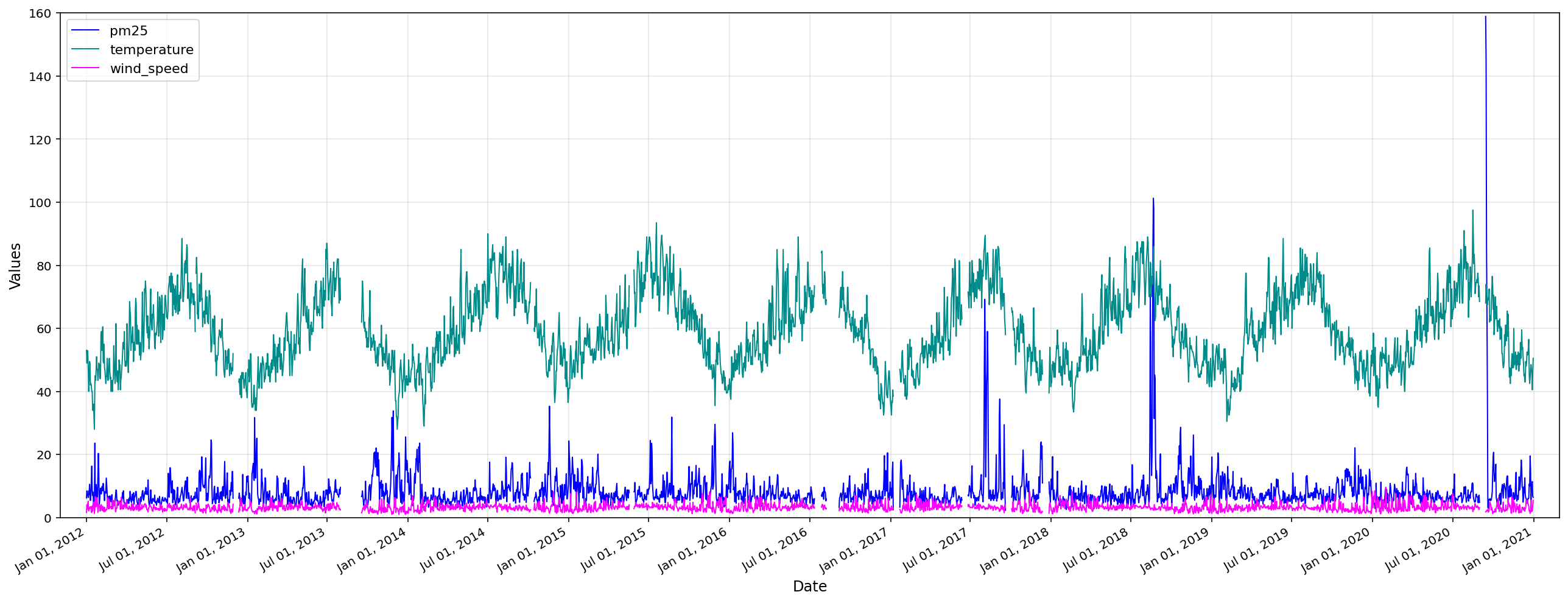}
\caption{Visualization of a multivariate time series: pm2.5 is the data to forecast, and wind speed and temperature are covariates to pm2.5.}\label{fig-arimax-example}
\end{figure}

The multivariate time series we use is visualized in Figure \ref{fig-arimax-example}. It is a daily time series, and has three components: 1) the average PM2.5 value; 2)the average wind speed; 3) the highest temperature. The PM2.5 data is the one to forecast, and wind speed and temperature are covariates. The historical time series spans from 2012-01-01 to 2020-12-31, and the forecast horizon is 30.

The evaluation results comparing ARIMA\_PLUS and ARIMA\_PLUS\_XREG are shown in Table \ref{arima-plus-xreg-result}, where we can see that ARIMA\_PLUS\_XREG generates more accurate forecasting results compared to ARIMA\_PLUS. These results can be reproduced by using this notebook which is published in Github \cite{xreg-notebook}.

\begin{table}[H]
\centering
\begin{tabular}{|c|c|c|c|c|c|}
\hline
 & MAE & MSE & RMSE & MAPE & sMAPE \\
\hline
ARIMA\_PLUS & 2.52 & 11.93 & 3.45 & 27.50 & 27.23 \\ \hline
ARIMA\_PLUS\_XREG & 2.33 & 9.04 & 3.01 & 26.79 & 25.34 \\ \hline
Accuracy improvement & 8.2\% & 32.0\% & 14.6\% & 2.7\% & 7.5\%\\
\hline
\end{tabular}
\caption{Evaluation results for ARIMA\_PLUS (univariate) and ARIMA\_PLUS\_XREG (multivariate). 
}\label{arima-plus-xreg-result}
\end{table}

\subsection{Forecasting Scalability}
We train the ARIMA\_PLUS model on the BigQuery public dataset 
``\href{https://console.cloud.google.com/bigquery?p=bigquery-public-data&d=iowa_liquor_sales&t=sales&page=table}{bigquery-public-data.iowa\_liquor\_sales}". We forecast liquor sales for a high volume of liquor products in different stores, between the date of 2015-01-01 and the date of 2021-12-31. We also duplicate the data to test the scalability for different data size up to 100 million time series. The model training uses a smaller context window ($max\_time\_series\_length = 32$) for speed-up, which doesn't much sacrifice forecasting accuracy according to the experiments in Section \ref{evaluate-hyperparameters}. The experiment utilized a dedicated BigQuery reservation with an allocation of 5,000 slots. A BigQuery slot is defined as the virtual compute unit employed by BigQuery to execute SQL queries.

\begin{figure}[H]
\centering
\includegraphics[width=0.95\textwidth]{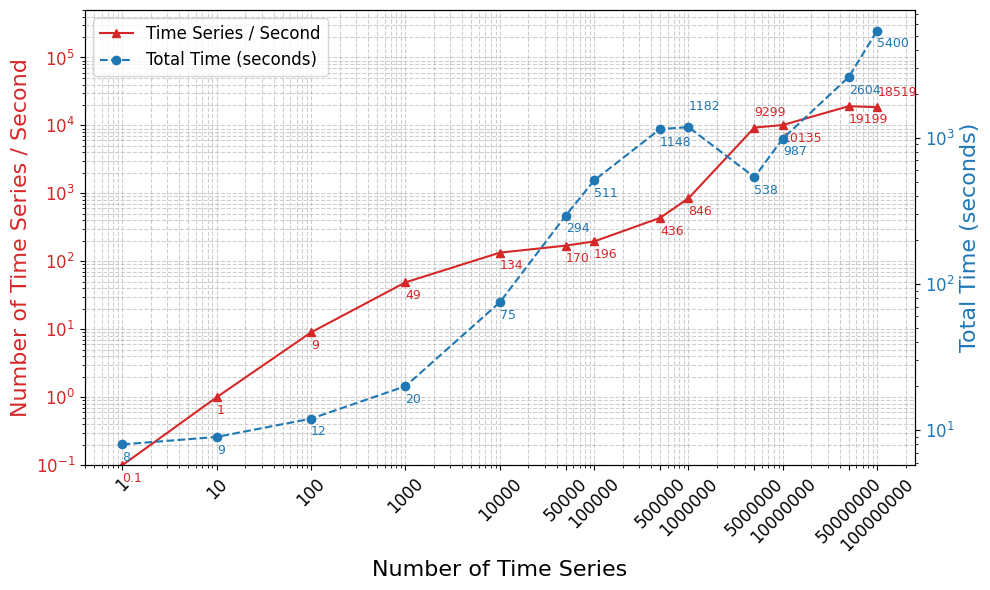}
\caption{The production throughput of $\text{ARIMA\_PLUS}$, measured as the total training time in seconds and the number of time series handled per second, shown across different dataset sizes. Counter-intuitively, the processing time for 5 million time series is shorter than for 1 million. This efficiency is due to improved data parallelism: the increased number of time series yields more data partitions, enabling a greater number of machines to process the workload simultaneously.
}\label{fig-scalability}
\end{figure}

We report the running time for different data sizes, and also calculate the number of processed time series per second for each data size. The results are illustrated in Figure \ref{fig-scalability}. From the figure, we can see that when the data size increases, BigQuery can achieve higher throughput thanks to resource auto-scaling. This allows forecasting 100 million time series using only 1.5 hours. 

\subsection{Forecasting Interpretability}
We use the following example to showcase the interpretability of ARIMA\_PLUS. These results can be reproduced by using this notebook published in Github \cite{custom-holiday-notebook}.

\begin{figure}[H]
\centering
\includegraphics[width=0.95\textwidth]{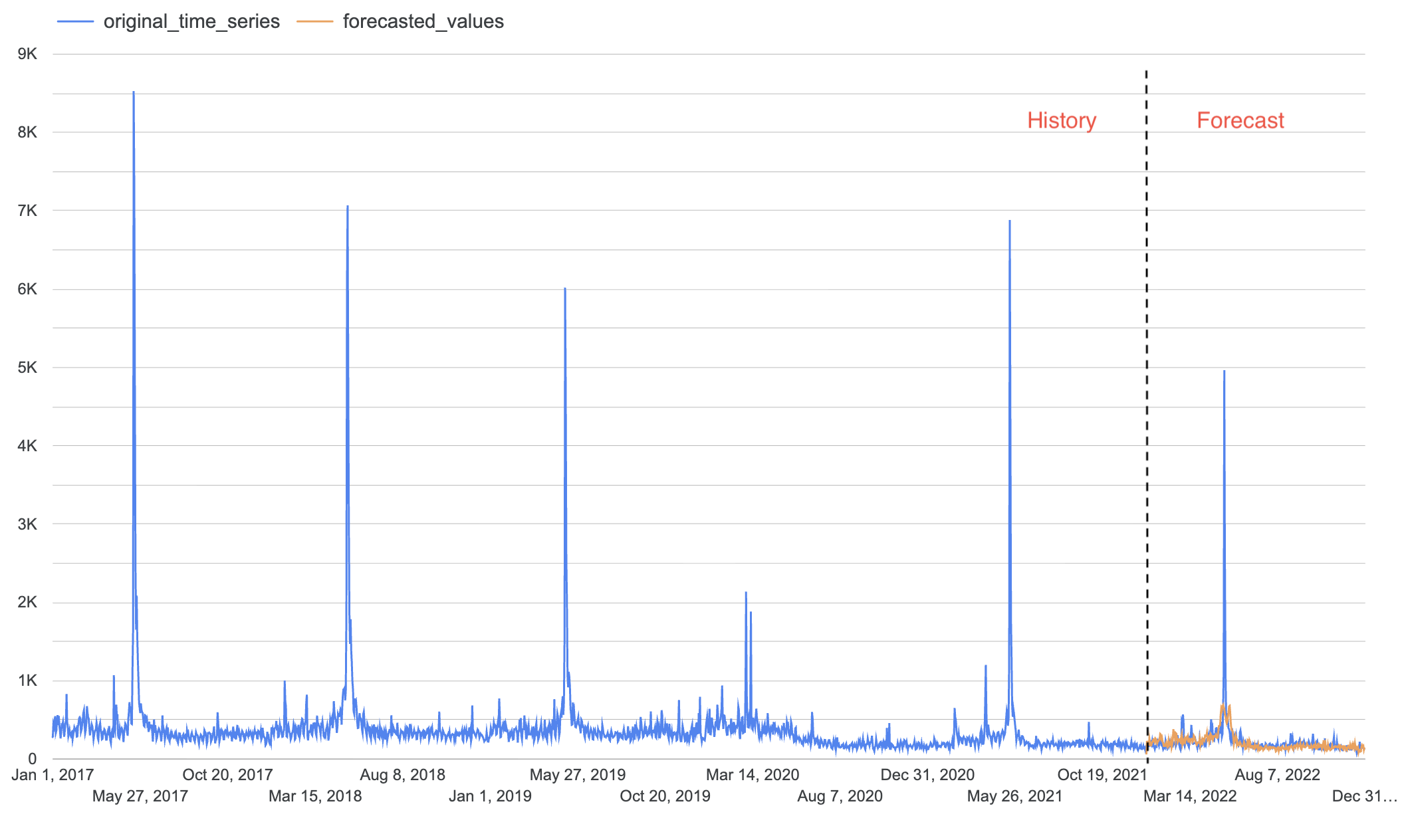}
\caption{Forecasting the Google I/O page views with the built-in US holidays only. For the input time series, the data points between 2017 and 2021 are used to forecast 2022. The forecasted values are overlaying with the actual values for a visual comparison.
}\label{fig-interpretability-without-custom-event}
\end{figure}

The  time series is the daily number of views for the Wikipedia ``Google I/O" page, from 2017-01-01 to 2021-12-31. First, we train an ARIMA\_PLUS model with built-in US holiday modeling with the holiday information available in the public dataset ``\url{bigquery-public-data.ml\_datasets.holidays\_and\_events\_for\_forecasting}", and forecast the next 365 days of the input time series, and visualize the result against the actual values in Figure \ref{fig-interpretability-without-custom-event}. As shown in the figure, the forecasting model captures the general trend and seasonalities pretty well. However, it doesn't capture the traffic spike in May 2022. In more detail, the generated forecasts have a small lift in that period but the level remains far below the actual value. 

\begin{figure}[H]
\centering
\includegraphics[width=0.95\textwidth]{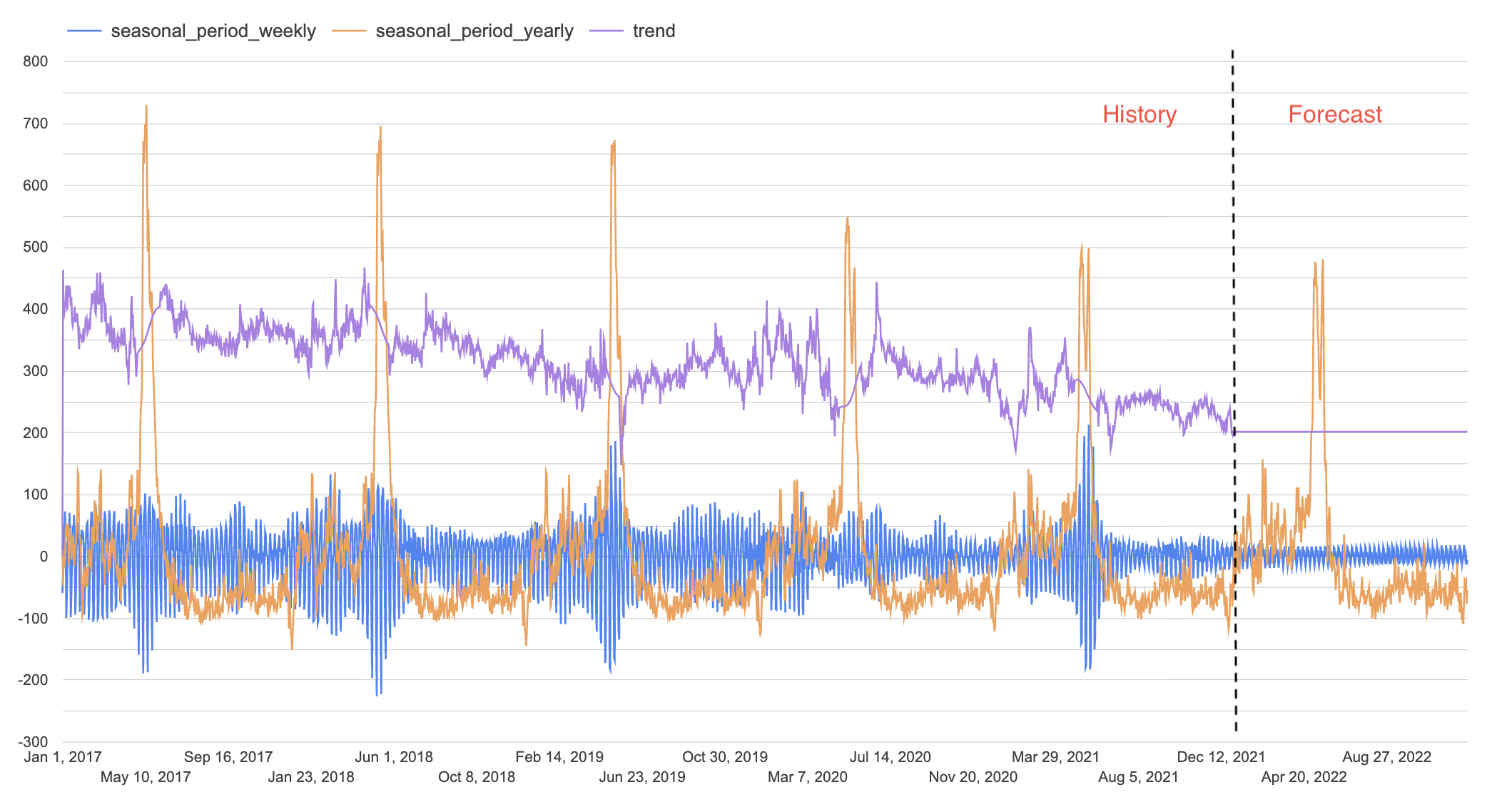}
\caption{Forecasting the Google I/O page views using ARIMA\_PLUS  with the built-in US holidays only. The trend and seasonal components identified by the ARIMA\_PLUS model are visualized. The data points for 2022 are forecasted values for these time series components.
}
\label{fig-interpretability-trend-seasonal}
\end{figure}

To understand and diagnose the above forecasting results, we visualize all the time series components that ARIMA\_PLUS detects in the historical time series and their corresponding forecasts. Figure \ref{fig-interpretability-trend-seasonal} shows that the final 2022 forecast consists of three components, namely the trend component, and two detected additive seasonal components (weekly and yearly). The historical fitting of each components are also visualized. From the figure, we can see that the trend forecast is flat according to the model, mostly because the `downward’ trend itself slowed down with the years. More interestingly, a consistent seasonal hump is extracted during May around Google I/O. Looking back in Figure \ref{fig-interpretability-without-custom-event}, we can see that in addition to the spikes at the exact dates of the events, there are some ramp up and ramp down effects that contribute a lot to the May seasonality. The seasonality module successfully detects and predicts such ramp up and ramp down effects, whose levels are indeed lower than the spikes at event dates. In Figure \ref{fig-interpretability-trend-seasonal}, we can also see that the historical magnitude of such ramp up/down effects becomes smaller throughout the years, and the seasonality module correctly extrapolates the size in 2022 as well.

\begin{figure}[H]
\centering
\includegraphics[width=0.95\textwidth]{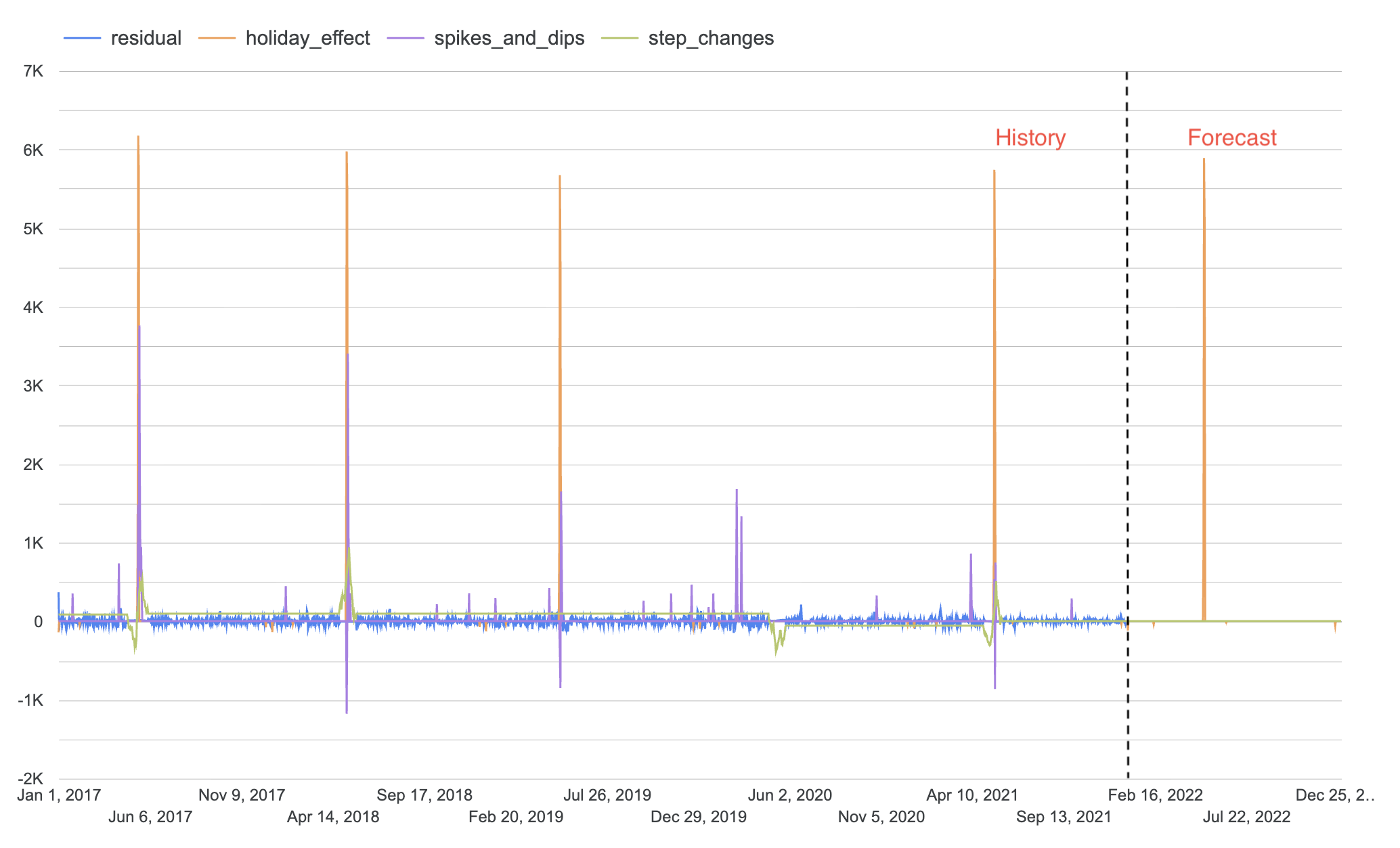}
\caption{Forecasting the Google I/O page views using ARIMA\_PLUS with the built-in US holidays only. The holiday effect, spikes and dips, step change and residual components identified by the ARIMA\_PLUS model are visualized. Only the holiday effect component has forecasted values, starting from 2022-01-01.
}\label{fig-interpretability-holidays}
\end{figure}

The detected holiday effects, spikes and dips, step change adjustments and the residual components are visualized in Figure \ref{fig-interpretability-holidays}. From the figure, we can see that the spikes at the Google I/O event dates in the input time series are detected as anomalies and removed for most of its values, even adjusting correctly for the unusual pattern of 2020 Google I/O during COVID. Some general change points that capture variation of traffic patterns throughout the years are also removed by the change point detection and step change adjustment module. The remaining parts are modeled via the yearly seasonal component, which generates the aforementioned hump in May of 2022. This automatic spike detection completely removes the outliers, and therefore improves the estimation of all other time series components, such as trend and seasonality, allowing them to emerge clearly rather than being muted or biased by outliers. Failing to detect these spikes results in a less meaningful decomposition, which is uninformative for understanding the historical context of the series, let alone achieving accurate forecasting.

We also demonstrate the interpretability of ARIMA\_PLUS using another example in \ref{google-360}. It forecasts the daily visits from Google Analytics 360 data and shows a more significant and meaningful step change adjustment than the Google I/O example, demonstrating the power of the change point detection and step change adjustment module.

\subsection{Forecasting Customization}
We continue to use the above example to showcase how to customize ARIMA\_PLUS to improve the forecasting results with additional signals.

\begin{figure}[H]
\centering
\includegraphics[width=0.95\textwidth]{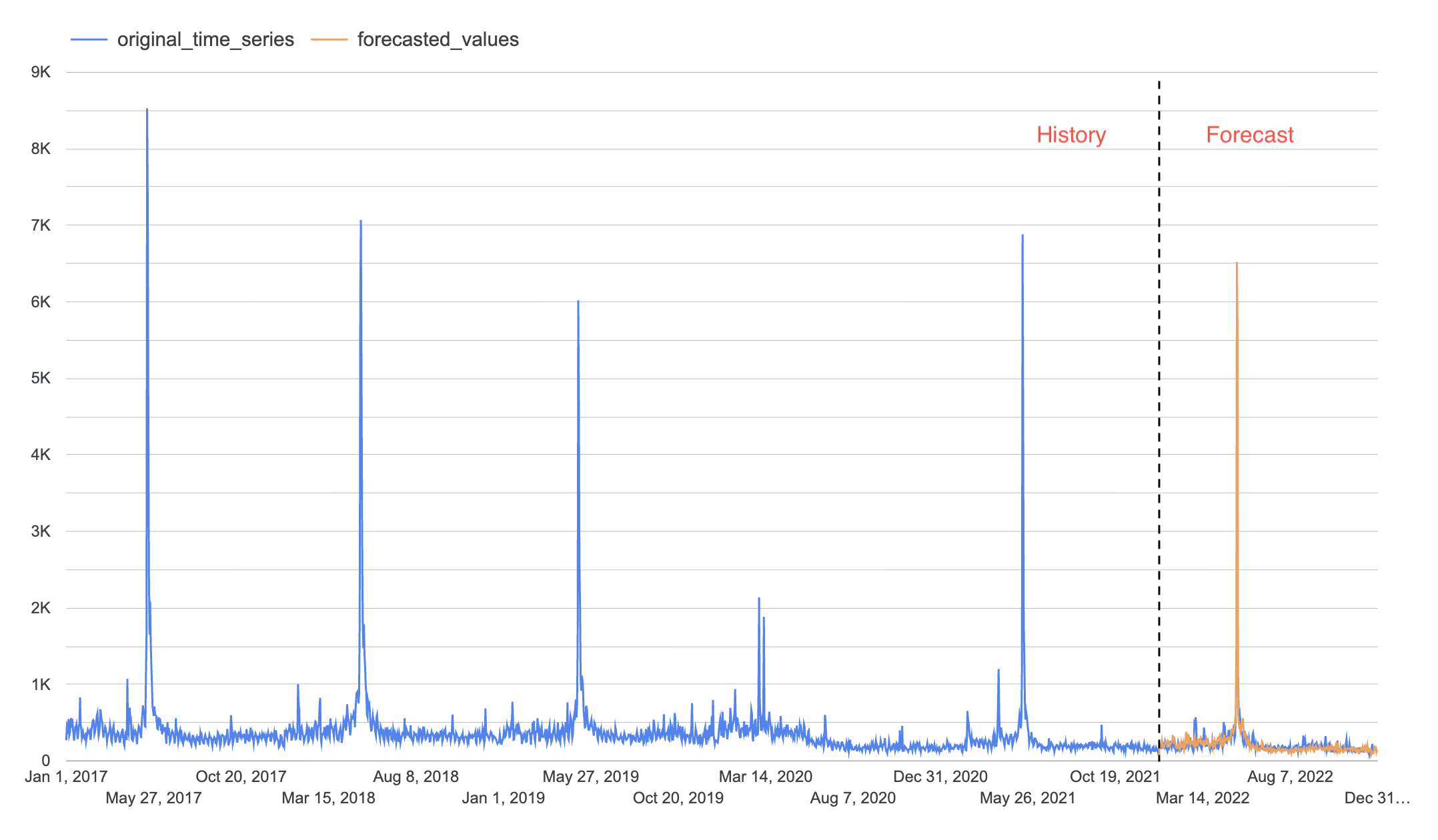}
\caption{Forecasting the Google I/O page views using ARIMA\_PLUS with the customized event dates for Google I/O, in addition to the built-in US holidays. The forecasted values are overlaying with the actual values for a visual comparison.
}\label{fig-interpretability-with-custom-events}
\end{figure}

For practitioners who are forecasting this time series, they should have prior knowledge that those spikes are around Google I/O events in the past years, and the event dates are all public, i.e., 2017-05-17, 2018-05-08, 2019-05-07, 2021-05-18 and 2022-05-11, and it was cancelled in 2020 due to pandemic. We can further incorporate these event days into the model training process, via the 'custom\_holiday' option. The model extracts previous lifts during Google I/O dates and applies smoothed effects to the future events. See section \ref{sec:holiday} for more details. The new forecasted results after supplying the additional event information are shown in Figure \ref{fig-interpretability-with-custom-events}, where they now effectively captures the increase of page views caused by the Google I/O event in 2022. We also evaluate the forecasting accuracy for the three days around the event (2022-05-08 to 2022-05-10), with and without holiday/event customization. The results are shown in Table \ref{custom-holiday-eval-result} where we can see that the forecasting accuracy has been significantly improved with the holiday/event customization. 

\begin{table}[H]
\centering
\begin{tabular}{|c|c|c|c|c|c|}
\hline
 & MAE & MSE & RMSE & MAPE & sMAPE \\
\hline
Without event customization & 1302.9 & 4711596.0 & 2170.6 & 66.0 & 73.1 \\ \hline
With event customization & 547.6 & 669358.3 & 818.1 & 34.8 & 27.7 \\ \hline
Accuracy improvement & 58.0\% & 85.8\% & 62.3\% & 47.3\% & 62.1\%\\
\hline
\end{tabular}
\caption{Forecasting evaluation results for ARIMA\_PLUS with and without customized holiday/events, for the three days around the Google I/0 event in 2022. 
}\label{custom-holiday-eval-result}
\end{table}

\begin{figure}[H]
\centering
\includegraphics[width=0.95\textwidth]{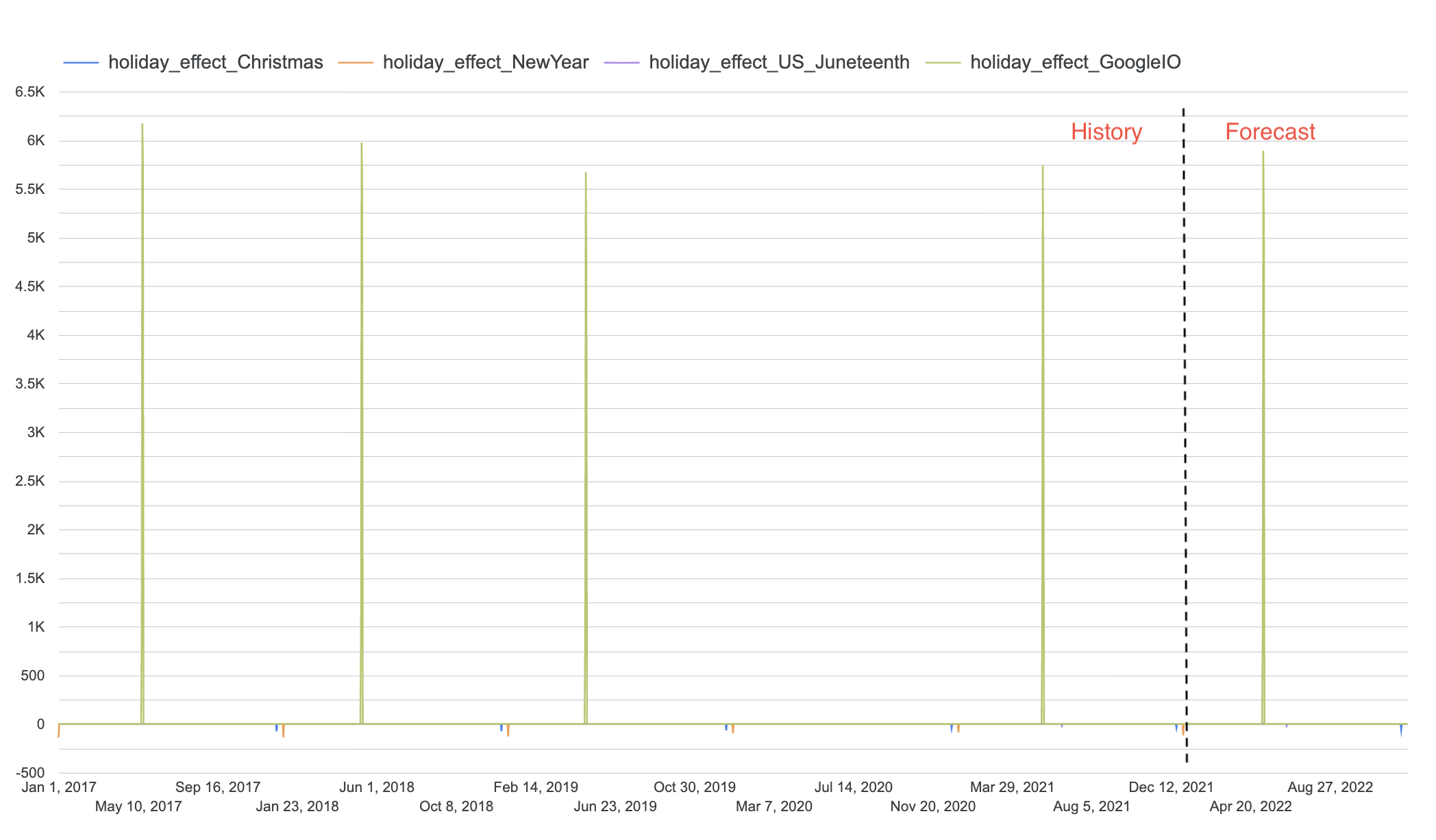}
\caption{Forecasting the Google I/O page views using ARIMA\_PLUS with the customized event dates for Google I/O, in addition to the built-in US holidays. The specific effect of each holiday and event identified by the $\text{ARIMA\_PLUS}$ model is visualized.
}\label{fig-holiday-breakdown}
\end{figure}

We further plot the per-holiday and per-event time series components in Figure \ref{fig-holiday-breakdown}, where we can see that in addition to the built-in holidays (Christmas, New year, Juneteenth), now the Google I/O event component is also captured.

We further quantify the percentile magnitude for every detected component in the historical time series, providing versions with and without customized Google I/O events. This magnitude is defined as the sum of squared values across all historical time points for each time series component. The results, shown in Table \ref{signal-magnitude}, demonstrate a significant shift thanks to the correct categorization of spikes: the magnitude of the holiday effect increases substantially (from 2.3\% to 50.9\%), while the contribution of spikes and dips drops sharply (from 51.4\% to 10.1\%). Correspondingly, the magnitude of the trend component also decreases (from 37.1\% to 28.5\%).

\begin{table}[H]
\centering
\begin{tabular}{|c|c|c|}
\hline
& without holiday customization & with holiday customization \\ \hline
trend & 37.1\% & 28.5\%\\ \hline
yearly seasonality & 5.9\% & 4.6\% \\ \hline
weekly seasonality & 1.0\% & 0.8\% \\ \hline
holiday effect & 2.3\% & 50.9\%\\ \hline
spikes and dips & 51.4\% & 10.1\% \\ \hline
step changes & 4.0\% & 4.6\% \\ \hline
residual & 0.5\% & 0.5\% \\ \hline
\end{tabular}
\caption{The percentile magnitude of each detected component in the historical time series, with and without customized holiday/events modeling. 
}\label{signal-magnitude}
\end{table}



\section{Conclusion}
\label{conclusions}
This work presents an innovative model and infrastructure framework, that enables automatic, scalable, accurate, and interpretable time series forecasting and anomaly detection, and establishes a new state-of-the-art for industry-scale applications. In terms of modeling, we propose a novel modular flow architecture that incorporates diverse options with unique enhancements. This architecture represents a class of models that processes a combination of time series components, including trend, seasonality, holiday effects, and exogenous covariates, and offers essential interpretability and customizability to business use cases. Furthermore, it follows a principled approach towards proper modeling of time series issues, and integrates specialized techniques for irregular seasonality, detection and handling of change points, trend model selections, etc., leading to improvements in accuracy without sacrificing interpretability. The benchmark on the 42 public datasets in the Monash forecasting repository over hundreds of thousands of diverse time series demonstrates that the proposed model achieves higher accuracy than a variety of well-established time series forecasting models across statistical and neural network alternatives. From a system perspective, the framework is implemented as a system, accessible via a simple SQL interface. Leveraging the managed cloud infrastructure, the system automatically scales to train independent models for 100 million time series in just 1.5 hours.

\section{Acknowledgements}
We thank Pablo Montero Manso and Sam McVeety for helpful reviews of this paper. We thank Skander Hannachi, Lisa Yin, and Mike Schachter for suggestions and insights throughout the development of ARIMA\_PLUS. We also thank Rajat Sen, Yichen Zhou, and Abhimanyu Das for their valuable discussion in benchmarking ARIMA\_PLUS against the Monash forecasting repository.




\appendix
\section{Forecasting Google Analytics 360 data using ARIMA\_PLUS}
\label{google-360}
\begin{figure}[H]
\centering
\includegraphics[width=0.95\textwidth]{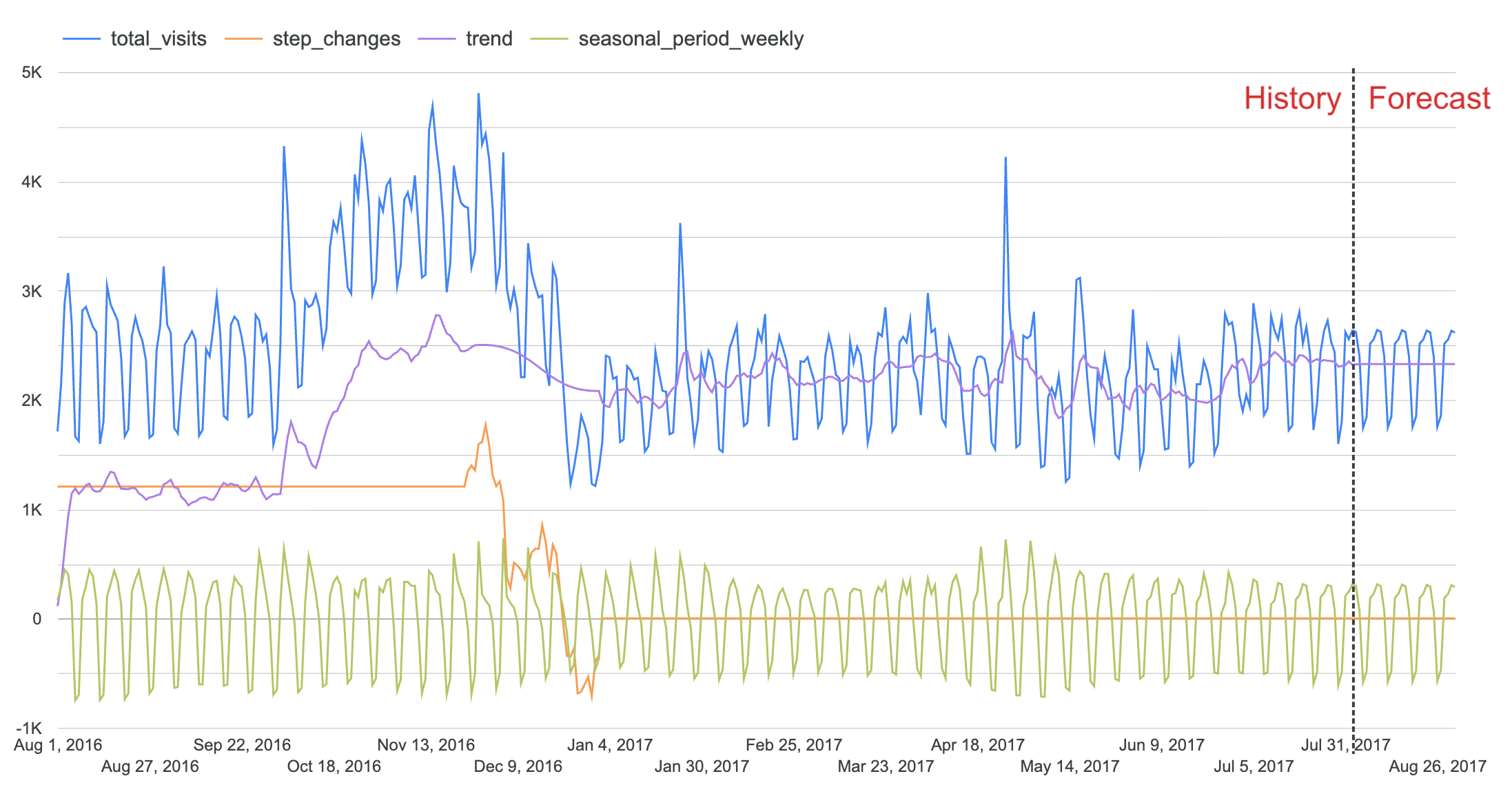}
\caption{Forecasting the daily number of visits of Google Analytics 360 data using ARIMA\_PLUS. The step changes, trend and seasonal components identified by the ARIMA\_PLUS model are visualized.
}
\label{fig-step-changes}
\end{figure}

We use the ``bigquery-public-data.google\_analytics\_sample.ga\_sessions" dataset to forecast the daily number of visits of the Google Analytics 360 data. The input time series spans from 2016-08-01 to 2017-07-31, with a forecast horizon of 31 days. The forecasting results are visualized in Figure \ref{fig-step-changes}. More details can be found in \url{https://docs.cloud.google.com/bigquery/docs/arima-single-time-series-forecasting-tutorial}.

\end{document}